\documentclass[aps, prd, twocolumn, amsmath, floats,floatfix, superscriptaddress, nofootinbib]{revtex4}

\usepackage{amssymb}
\usepackage{amsmath}
\usepackage{verbatim}
\usepackage{mathrsfs}
\usepackage{amsfonts}
\usepackage{latexsym}
\usepackage{epsfig}
\usepackage{color}
\usepackage{graphicx,subfigure}
\usepackage{units}
\usepackage{mathtools}
\usepackage{envmath}
\usepackage{natbib}
\usepackage{ctable}
\usepackage{soul}
\usepackage{ulem}
\usepackage{hyperref}

\begin{document}


\definecolor{orange}{rgb}{0.9,0.45,0}

\newcommand{\re}{\mbox{Re}}
\newcommand{\im}{\mbox{Im}}

\newcommand{\miq}[1]{\textcolor{orange}{#1}}
\newcommand{\red}[1]{{\color{red} #1}}
\newcommand{\pcd}[1]{{\color{blue} #1}}

\def\CovDev{D}
\def\Res{{\mathcal R}}
\def\Gammaflat{\hat \Gamma}
\def\metricflat{\hat \gamma}
\def\Dflat{\hat {\mathcal D}}
\def\part_n{\partial_\perp}

\def\Lie{\mathcal{L}}
\def\A{\mathcal{X}}
\def\Aphi{\A_{\phi}}
\def\hAphi{\hat{\A}_{\phi}}
\def\E{\mathcal{E}}
\def\Ham{\mathcal{H}}
\def\M{\mathcal{M}}
\def\R{\mathcal{R}}
\def\p{\partial}

\def\hg{\hat{\gamma}}
\def\hA{\hat{A}}
\def\hD{\hat{D}}
\def\hE{\hat{E}}
\def\hR{\hat{R}}
\def\hcA{\hat{\mathcal{A}}}
\def\hDelt{\hat{\triangle}}

\def\na{\nabla}
\def\dif{{\rm{d}}}
\def\non{\nonumber}
\newcommand{\erf}{\textrm{erf}}

\renewcommand{\t}{\times}

\long\def\symbolfootnote[#1]#2{\begingroup%
\def\thefootnote{\fnsymbol{footnote}}\footnote[#1]{#2}\endgroup}


\title{Prospects for the inference of inertial modes from hypermassive neutron stars with future gravitational-wave detectors}

\author{Miquel Miravet-Tenés}
\affiliation{Departament d'Astronomia i Astrofísica, Universitat de València, C/ Dr Moliner 50, 46100, Burjassot (València), Spain} 
\author{Florencia L. Castillo}
\affiliation{LAPP, Universit\'e Grenoble Alpes, Universit\'e Savoie Mont Blanc, CNRS/IN2P3, Annecy, France}
\author{Roberto De Pietri}
\affiliation{Dipartimento di Scienze Matematiche, Fisiche e Informatiche, Università di Parma, Parco Area delle Scienze 7/A,I-43124 Parma, Italy} 
\affiliation{INFN, Sezione di Milano Bicocca, Gruppo Collegato di Parma, I-43124 Parma, Italy}
\author{Pablo Cerdá-Durán}
\affiliation{Departament d'Astronomia i Astrofísica, Universitat de València, C/ Dr Moliner 50, 46100, Burjassot (València), Spain} 
\affiliation{ Observatori Astronòmic, Universitat de València, C/ Catedrático José Beltrán 2, 46980, Paterna (València), Spain}
\author{José A. Font}
\affiliation{Departament d'Astronomia i Astrofísica, Universitat de València, C/ Dr Moliner 50, 46100, Burjassot (València), Spain} 
\affiliation{ Observatori Astronòmic, Universitat de València, C/ Catedrático José Beltrán 2, 46980, Paterna (València), Spain}


\date{\today}

\begin{abstract}
Some recent, long-term numerical simulations of binary neutron star mergers have shown that the long-lived remnants produced in such mergers might be affected by convective instabilities. Those would trigger the excitation of inertial modes, providing a potential method to improve our understanding of the rotational and thermal properties of neutron stars through the analysis of the modes' imprint in the late post-merger gravitational-wave signal. 
In this paper we assess the detectability of those modes by injecting numerically generated post-merger waveforms into colored Gaussian noise of second-generation and future detectors. Signals are recovered using \textsc{BayesWave}, a Bayesian data-analysis algorithm that reconstructs them through a morphology-independent approach using series of sine-Gaussian wavelets.
Our study reveals that current interferometers (i.e.~the Handford-Livingston-Virgo network) recover the peak frequency of inertial modes only if the merger occurs at distances of up to 1 Mpc. For future detectors such as the Einstein Telescope, the range of detection increases by about a factor 10.
\end{abstract}

\maketitle

\section{\label{sec:sec1}Introduction}

Binary neutron star (BNS) mergers are among the most important sources of gravitational waves (GWs). Their detection offers the opportunity to improve our understanding of the physics of neutron stars (NS) and, in particular, constrain the equation of state (EOS) of such compact objects at supranuclear densities. So far, the LIGO-Virgo-KAGRA (LVK) \cite{LIGO:2015,Virgo:2015,KAGRA:2019} Collaboration has reported the observation of GWs from two such mergers, GW170817~\cite{Abbott:2017_GW170817} and GW190425~\cite{Abbott:2020}. The former event also produced an electromagnetic (EM) counterpart, GRB170817A/AT2017gfo~\cite{grb,Abbott:2017b}, which initiated the long-anticipated field of multi-messenger astrophysics with GWs~\cite{Abbott:2017b,Abbott:2017c}. The EM signature of GW170817 indicates the presence of an optical transient known as a kilonova~\cite{Pian:2017,Kasen:2017,Cowperthwaite:2017}, providing convincing support to the theoretical claim that identifies BNS mergers as likely progenitors of short gamma-ray bursts (sGRBs)~\cite{Kouveliotou:1993,MacFadyen:1999}. 

The investigation of the dynamics of BNS mergers, their post-merger evolution, and their GW emission strongly relies on numerical-relativity simulations. This field has undergone major advances during the last few years  (see~\cite{Baiotti:2017,Dietrich:2018,Duez:2019,Shibata:2019,Ciolfi:2020,Ruiz:2021,Sarin:2021} and references therein). Depending on the initial conditions of the binary system, mainly its total mass and the choice of EOS, a likely outcome of a BNS merger is a spinning black hole (BH) surrounded by an accretion disk. Momentarily, in a timescale of tens of milliseconds and before the BH forms, the post-merger object
can be a hypermassive neutron star (HMNS)~\cite{Baumgarte:2000}. This is the expected outcome when the total mass of the system is larger than the maximum mass of a cold, uniformly rotating NS (see~\cite{Akmal:1998} where the maximum mass for a large sample of cold EOS determined by solving the Tolman-Oppenheimer-Volkoff equation is shown to be in the range of $1.8-2.3 M_{\odot}$). A HMNS is supported against gravitational collapse by both differential rotation and thermal pressure. This transient object will ultimately collapse to a BH once its support against gravity lessens due to the loss of angular momentum to GW emission and dissipative effects~\cite{Shibata:2019,Ciolfi:2020,Ruiz:2021,Sarin:2021}.

During the first few milliseconds after its formation, the HMNS exhibits strong non-axisymmetric deformations and nonlinear oscillations, namely combinations of oscillation modes and spiral deformations~\cite{Stergioulas:2011,Hotokezaka:2013,Bauswein:2015,Takami:2015,Bauswein:2016,Bauswein:2019}. This is accompanied by the emission of GWs in a range of frequencies around a few kHz~\cite{Shibata:2000,Oechslin:2002,Baiotti:2008,Stergioulas:2011,Bauswein:2012,Lehner:2016,Rezzolla:2016,DePietri:2016,Dietrich:2017}. 
The GW spectrum of the HMNS is characterized by the presence of many distinct peaks (see e.g.~\cite{Bauswein:2019} for a review). The detection and interpretation of post-merger GW signals relies on a proper understanding of the physical mechanisms generating those  features in the spectrum. Through their analysis, inference on NS properties might be possible. In particular,  information on the EOS of the remnant star can be obtained through the study of the frequency of the $m=2$ \textit{f}-mode (quadrupolar mode) \cite{Shibata:2005,Kastaun:2010,Kastaun:2015,Clark:2016,Kastaun:2016,Kastaun:2017,Lioutas:2021,Soultanis:2022,Iosif:2022,Wijngaarden:2022}. There exists a significant amount of work to build empirical relations to infer the NS radius from the frequency peak ($f_{\rm peak}$) of this dominant mode~\cite{Bauswein:2012b,Chatz:2017,Bose:2018,Bauswein:2019}. The frequency peaks of the post-merger spectra can also be related to other NS properties, such as the tidal coupling constant~\cite{Bernuzzi:2015} or the average density~\cite{Takami:2015}. The empirical relations that link the GW spectrum and physical quantities of the HMNS can directly constrain the EOS (see~\cite{Takami:2015,Bauswein:2019} and references therein). 

On timescales longer than about 50 ms after merger the simulations of~\cite{DePietri:2018,DePietri:2020} (see also~\cite{Ciolfi:2019}) have shown the appearance and growth of convective instabilities in the remnant. The simulations, based on a piecewise polytropic approximation for the EOS treatment supplemented by a thermal component~\cite{Read:2009}, showed that at $40-50$ ms after merger (depending on the EOS), the amplitude of the $m=2$ \textit{f}-mode, which is the dominant mode in the early and intermediate post-merger phases, has noticeably decreased. By that time, convective instabilities set in and trigger inertial modes. The GW emission associated with those modes is found to dominate over the initial $m= 2$ \textit{f}-mode at late post-merger times, producing new distinctive peaks in the HMNS GW spectrum. The post-merger timescales discussed in~\cite{DePietri:2018,DePietri:2020} at which the HMNS is affected by convective instabilities  are compatible with those found by~\cite{Camelio:2019} who analyzed convectively unstable rotating NS with non-barotropic thermal profiles (as in the case of BNS remnants). Since inertial modes depend on the rotation rate of the star and they are triggered by convection, their detection in GWs would provide a unique opportunity to probe the rotational and thermal state of the merger remnant. As an example to conduct such inference, an empirical relation between the frequency of the inertial modes and the angular velocity and the rotation rate of the star was proposed by~\cite{Kastaun:2008}.

The results of~\cite{DePietri:2018,DePietri:2020} indicate that the GW emission of inertial modes in the late post-merger phase is potentially detectable by the planned third-generation GW detectors. In this paper we further investigate this issue by reconstructing the GW signals  of~\cite{DePietri:2020} using \textsc{BayesWave}\footnote{\href{https://git.ligo.org/lscsoft/bayeswave}{https://git.ligo.org/lscsoft/bayeswave}} \cite{Cornish:2015,Littenberg:2015}, a Bayesian data-analysis algorithm that recovers the post-merger signal through a morphology-independent approach using series of sine-Gaussian wavelets. To assess the detectability of inertial modes we perform injections into the noise of different detectors from sources at different distances: the current Hanford-Livingston-Virgo (HLV) detector network \cite{Harry:2010,LALSuite, Virgo:2015} and the future Einstein Telescope (ET)~\cite{ET:2010, Hild:2011}. We also check the dependence of our results on the NS EOS by using two different EOS, APR4 and SLy ~\cite{Read:2009}. The reconstructed waveform distributions that we obtain for each injection allows us to infer posteriors of the peak inertial mode frequency, $f_{\rm inertial}$. Our analysis shows that inertial modes can be potentially detected by third-generation GW detectors up to distances of about $10$ Mpc. 

The paper is organized as follows: in Section~\ref{sec:reconstruction} we briefly present the \textsc{BayesWave} algorithm and introduce the quantities we use to assess the waveform reconstructions. Our main results are presented in Sec.~\ref{sec:results} where we briefly describe the numerical-relativity simulations used to generate the waveforms employed for the injections and we discuss the waveform reconstruction performance.
Finally, our conclusions are presented in  Section~\ref{sec:discussion}. 

\section{\label{sec:reconstruction}Waveform reconstruction}

\subsection{\label{sec:BAYESWAVE} The \textsc{BayesWave} algorithm}

The goal of this work is to analyze the reconstruction of the GW signal produced after the merger of two NS, particularly in the late post-merger phase. To do so we employ \textsc{BayesWave}, a Bayesian signal reconstruction algorithm that uses Morlet-Gabor (or sine-Gaussian) wavelets  \cite{Cornish:2015,Littenberg:2015} to model morphologically unknown non-Gaussian features with minimal assumptions \cite{BECSY:2017}. In the time domain, the two  GW polarizations of the wavelets are given by
\begin{align}
    h_+(t) & = A e^{-(t-t_0)^2/\tau^2}\cos{[2\pi f_0(t-t_0)+\phi_0]}\,, \\
    h_{\times}(t) & = \epsilon h_+(t)e^{i\pi/2}\,, \label{e_eq}
\end{align}
where $A$ is the amplitude of the wavelet, $f_0$ is the central frequency, $t_0$ is the central time, $\phi_0$ is the offset phase, $\epsilon$ is the ellipticity, and $\tau = Q/(2\pi f_0) $, where $Q$ is the quality factor~\cite{Cornish:2015}. 
The factor $e^{i\pi/2}$ in Eq.~\eqref{e_eq} indicates there is a $\pi/2$ difference in the phase of both polarizations.

\textsc{BayesWave} employs a trans-dimensional reversible jump Markov chain Monte Carlo (RJMCMC) to sample the joint posterior of the parameters of the wavelets, the number $N_W$ of wavelets and ellipticity. These are used to derive the posterior {distribution} of the reconstructed waveform and, using the waveform samples, it is straightforward to obtain posteriors of quantities that can be derived from the signal. This sampler ensures that the algorithm does not overfit the data, since the addition of wavelets to the reconstruction increases the dimensionality of the model, which provokes a reduction of the posterior probability. There has to be a balance between the improvement of the fit and the addition of wavelets in order to overcome the Occam penalty~{\citep{Smith:1980}}. 

\subsection{\label{sec:overlap_fpeak} Overlap and Peak Frequency}

A way to check how well a signal that is injected into detector noise is recovered is the use of the \textit{overlap} function between the injected signal, $h_i$, and the recovered model from \textsc{BayesWave}, $h_r$:
\begin{equation}\label{overlap}
    \mathcal{O} = \frac{\langle h_i, h_r \rangle}{\sqrt{\langle h_i,h_i\rangle}\sqrt{\langle h_r,h_r\rangle}}\,,
\end{equation}
where the inner product of two complex quantities $a$ and $b$, $\langle a,b\rangle$, is defined as
\begin{equation}\label{inner_prod}
    \langle a,b \rangle \equiv 2\int_0^{\infty}\frac{a(f)b^*(f)+a^*(f)b(f)}{S_h(f)}df\,,
\end{equation}
where $S_h(f)$ refers to the one-sided noise power spectral density of the detector and the asterisk denotes complex conjugation. The value of the overlap function ranges from -1 to 1, being $\mathcal{O} = 1$ a perfect match between the injected and the reconstructed signal, $\mathcal{O} =-1$ a perfect anti-correlation, and $\mathcal{O} = 0$ means no match between the signals. One can also compute the weighted overlap from a network of $N$ detectors:
\begin{equation}\label{net_overlap}
    \mathcal{O}_{\rm network} = \frac{\sum^N_{k=1}\langle h_i^{(k)}, h_r^{(k)} \rangle}{\sqrt{\sum^N_{k=1}\langle h_i^{(k)},h_i^{(k)}\rangle}\sqrt{\sum^N_{k=1}\langle h_r^{(k)},h_r^{(k)}\rangle}}\,,
\end{equation}
where the index $k$ stands for the $k$-th detector. 
With the resulting overlap between the injected and reconstructed signals from \textsc{BayesWave} we assess the reconstructions for different distances to the GW source (i.e.~different signal-to-noise ratios (SNRs)). 

We  also compute the peak frequency, defined as the one corresponding to the maximum value of the fast Fourier transform (FFT)~{\citep{Cooley:1965}} of the time-domain signal, $|\tilde{h}(f)|$, using a time window over the part of the signal we are interested in. The segments of data have been previously Hann-windowed. We expect the peak frequencies, $f_{\rm peak}$ and $f_{\rm inertial}$, to be located in the range $f\in [1500,4000]$ Hz \cite{Chatz:2017,DePietri:2020}, and we will use this range to set the low-frequency and high-frequency cut offs for the computation of the overlap and the frequency peaks.

\section{\label{sec:results}Results}

\subsection{\label{sec:simulations} Summary of the numerical-relativity simulations}

The waveforms we employ for our study were obtained in the  numerical-relativity simulations of BNS mergers performed by~\cite{DePietri:2020}. The initial data are generated using the \textsc{Lorene} code \cite{Gourgoulhon:2001,Gourgoulhon:2016} and the initial separation of the two stars is $\approx 44.3$ km, wich corresponds to about four full orbits before merger. The main properties of the initial simulation setup are reported in Table~\ref{tab:table_sim}. 
The evolution of the initial data is performed using the  \textsc{Einstein Toolkit}~\cite{Loffler:2012}, an open source code based on the Cactus framework \cite{Goodale:2003}. The simulation setup employed in the study of~\cite{DePietri:2020} is the same as in \cite{Maione:2017,DePietri:2016,DePietri:2019}, to which the reader is addressed for further details, except for the fact that $\pi$-symmetry was used to reduce the computational cost by a factor 2. The \textsc{Einstein Toolkit} solves Einstein's field equations in the BSSN formalism~\cite{Shibata:1995,Baumgarte:1998} and the general relativistic hydrodynamics equations in the Valencia formulation~\cite{Banyuls:1997,Font:2008}. The latter are integrated numerically with a finite-volume algorithm based on the HLLE Riemann solver \citep{Harten:1983,Einfeldt:1988}, the WENO reconstruction method \citep{Liu:1994,Jiang:1996}, and the Method of Lines with a $4^{\rm th}$ order, conservative Runge-Kutta scheme \citep{Shu:1988}. 

\begin{table}
    \centering
\begin{tabular}{cccccc}
\hline
    EOS & $M_0$ & $M$ & $C$ & $J_{\rm ADM}$ & $\Omega_0$ (krad/s) \\
    \hline 
    \hline
    APR4 & 1.4 & 1.2755 & 0.166 & 6.577 & 1.767\\
    SLy & 1.4 & 1.2810 & 0.161 & 6.623  & 1.770\\
\hline
\end{tabular}
    \caption{Main properties of our two BNS simulations. The columns report the EOS, the baryonic mass, $M_0$, the gravitational mass (at infinite distance), $M$, and the compactness, $C\coloneqq M/R$, of the individual stars, and the total angular momentum,$J_{\rm ADM}$, and the total angular velocity, $\Omega_0$, of the binary sistem. Geometrized  units are used ($c=G=M_{\odot} = 1 $).}
    \label{tab:table_sim}
\end{table}
 
The inertial modes identified in the simulations of~\cite{DePietri:2018,DePietri:2020} are triggered by a convective instability appearing in the non-isentropic HMNS which was identified by monitoring the value of the Schwarzschild discriminant. The modes have frequencies slightly smaller than twice the maximum angular frequency of the differentially rotating remnant star $\Omega_{\rm max}$. 

\subsection{\label{sec:rec_results}Waveform reconstruction performance}

\begin{figure*}[ht]
\centering
   \includegraphics[width=0.97\linewidth]{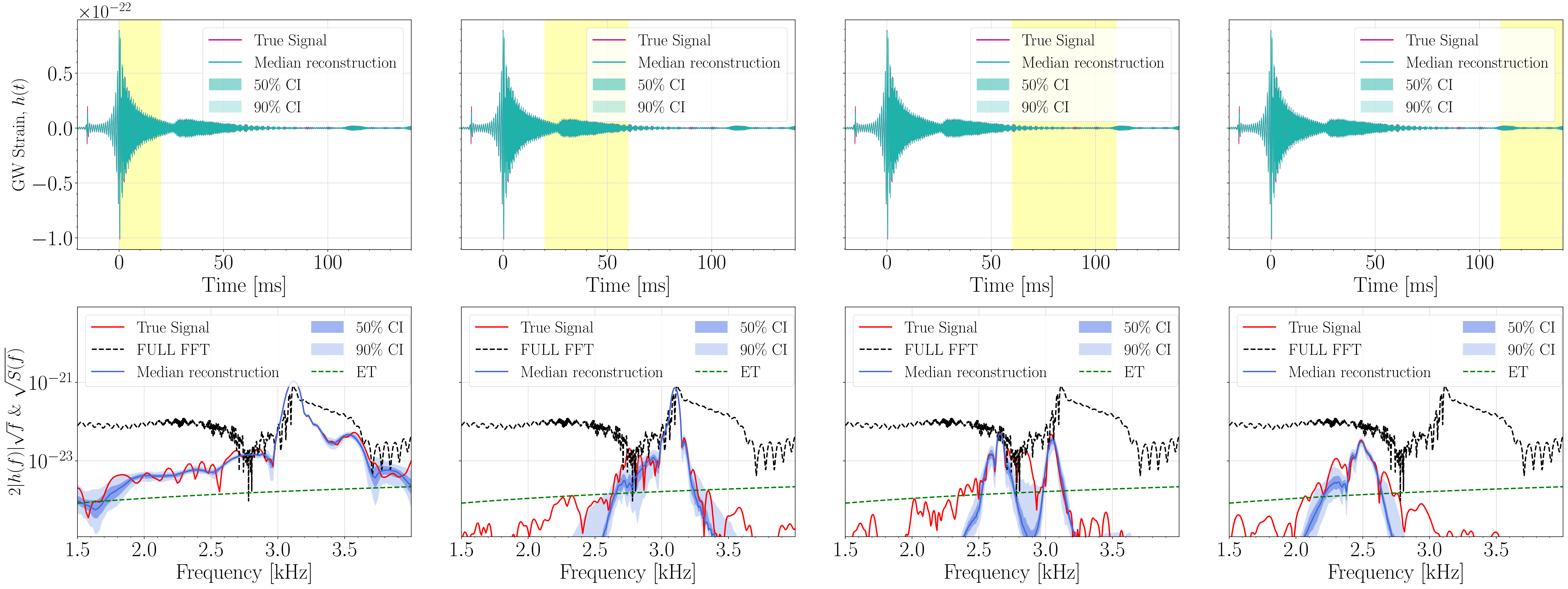}

\caption{Injected (red) and recovered (blue) time-domain waveforms (top panels) and ASD (bottom panels) for BNS merger simulations with the APR4 EOS. Each ASD is computed using the corresponding time window depicted in yellow in the top panels.  The source is assumed to be located at $d=3$ Mpc. The signals are injected into the E3 configuration of the third-generation ET observatory, whose sensitivity curve is shown by the dashed green curve. The width of the time windows is chosen to show  how the frequency peak is displaced to lower frequencies depending on the different evolutionary stage of the post-merger remnant.}
\label{fig:ETTimeWindow}
\end{figure*}

\begin{figure*}[ht]
\centering
   \includegraphics[width=0.97\linewidth]{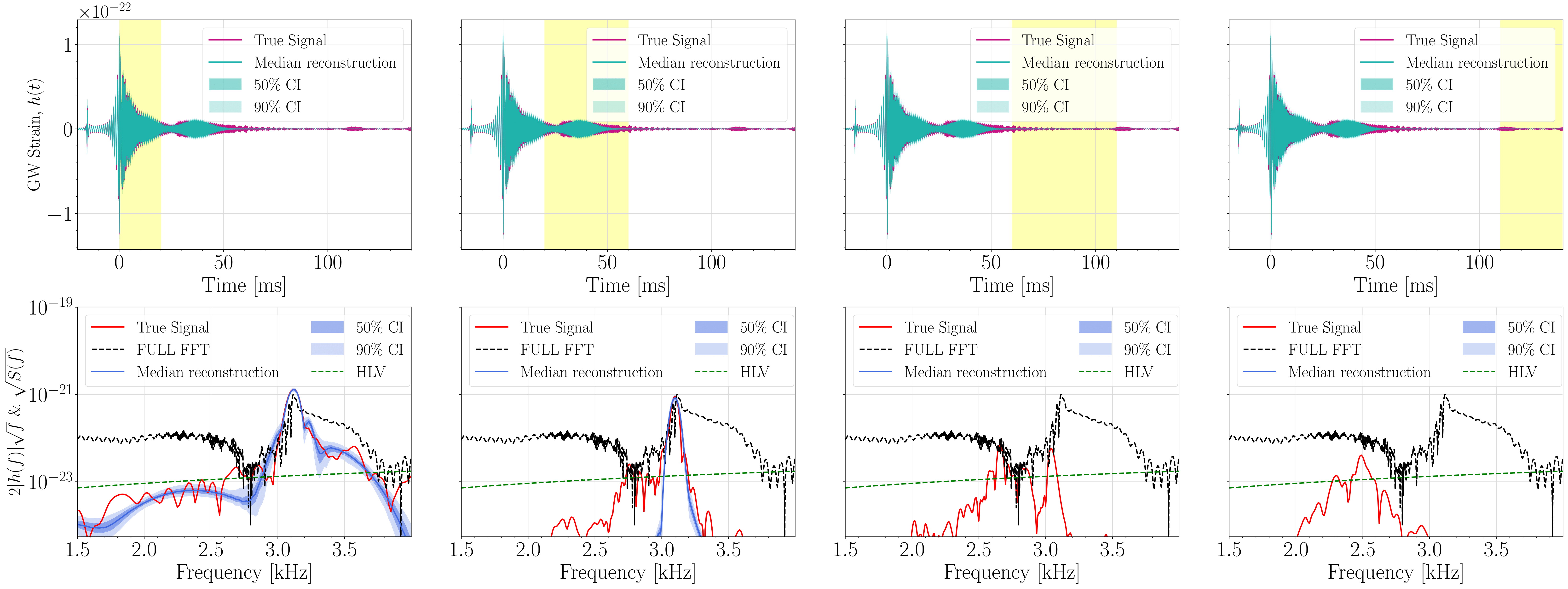}
\caption{Same as Fig.~\ref{fig:ETTimeWindow} but for the H1 detector.} 
\label{fig:H1TimeWindow}
\end{figure*}

In order to obtain a distribution of frequency peaks we perform injections in \textsc{BayesWave} of the waveforms computed by~\citep{DePietri:2020} . We use several sensitivity curves (for aLIGO we use the PSD model \texttt{aLIGOZeroDetHighPower} for the two detectors from \cite{LALSuite}, for Advanced Virgo we use the design sensitivity from \cite{Virgo:2015} and we take the ET-D configuration from \cite{Hild:2011}) to see the differences between current and future GW detectors. The reconstructions are compared using the design sensitivities of the HLV detector network and of the ET, formed by a three detector network on the same site. No sources of noise and/or glitches are added, we only consider Gaussian noise \citep{Blackburn:2008,Abbott:2009,Aasi:2012} colored by the PSD of the detector. We set the source of the injected signals at different distances (giving different SNRs) and assume that the source is also optimally oriented with respect to one of the detectors (Hanford, H1, for HLV and E3 for ET\footnote{The design of the Einstein Telescope consists of three arms forming an equilateral triangle, with three pairs of interferometers acting as a three-detector network, E1, E2 and E3.}). We set a maximum number of wavelets of $N_W^{\rm max} = 100$ for HLV and $N_W^{\rm max} = 200$ for ET, a maximum quality factor of {$Q^{\rm max} = 200$}, $n=2\times 10^6$ iterations, and a sampling rate of 8192 Hz.
The maximum number of wavelets is different for HLV and ET because selecting $N_W^{\rm max} = 100$ for ET is not large enough for the algorithm to reconstruct the signal accurately, due to the high sensitivity of third-generation detectors.

\begin{figure*}[ht]
\centering
   \includegraphics[width=0.95\linewidth]{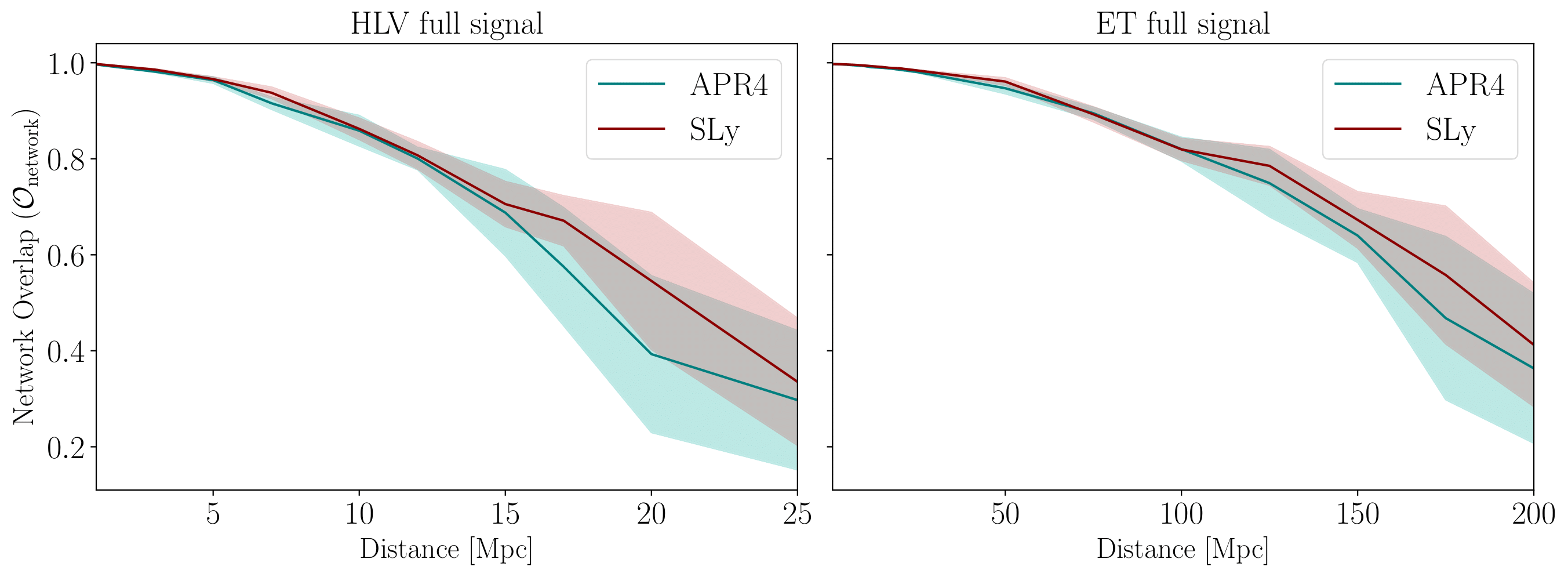}
\caption{Detector network overlap between the full injected and recovered signals as a function of the distance to the GW source. The left panel shows the results for the HLV network and the right one for the ET detector. The lines indicate the mean value over the waveform posterior distribution and the shaded areas are the standard deviations. }
\label{fig:overlap_full}
\end{figure*}

\begin{figure*}[ht]
\centering
   \includegraphics[width=0.95\linewidth]{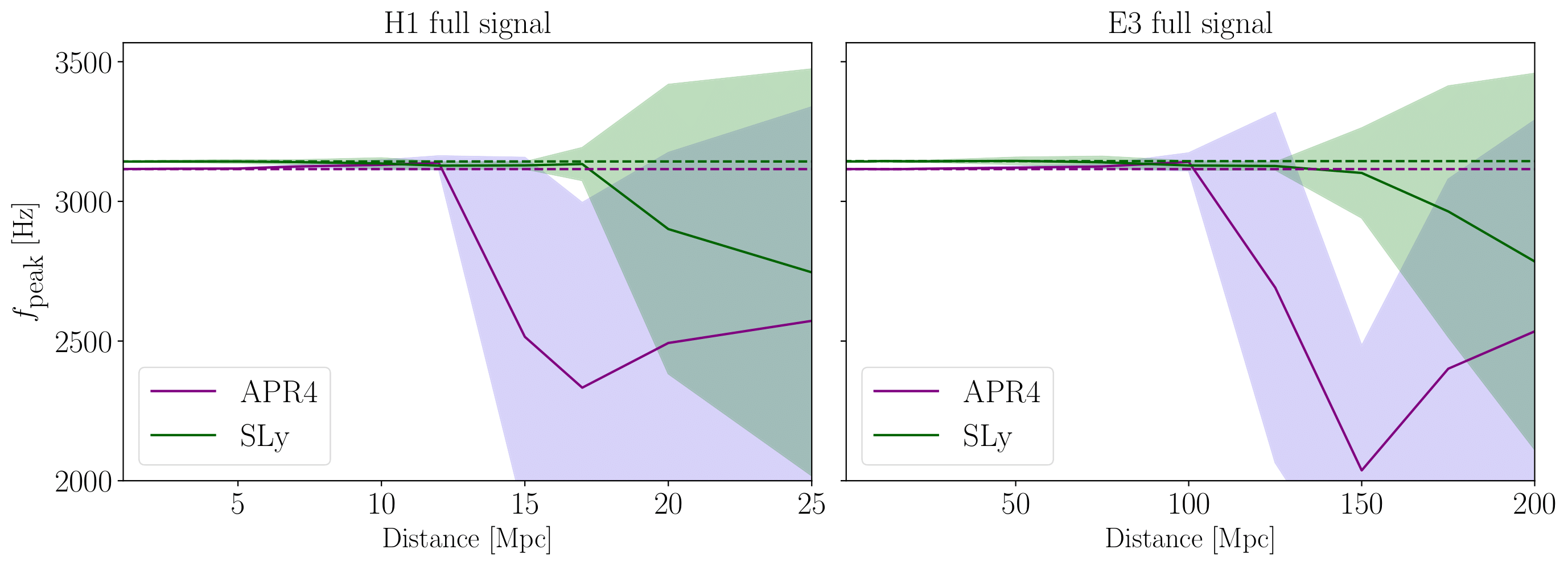}
\caption{Dependence of the recovered $f-$mode frequency peak with distance for the H1 detector (left panel) and for the E3 detector (right panel). The solid curves are the mean values and the shaded areas represent the standard deviations of the distributions. The mean of the recovered peak is close to the injected signal (dashed lines) for distances up to 10 Mpc for H1 and 100 Mpc for E3.}
\label{fig:fpeak_full}
\end{figure*}

\begin{figure*}[ht]
    \centering
    \includegraphics[width=0.95\textwidth]{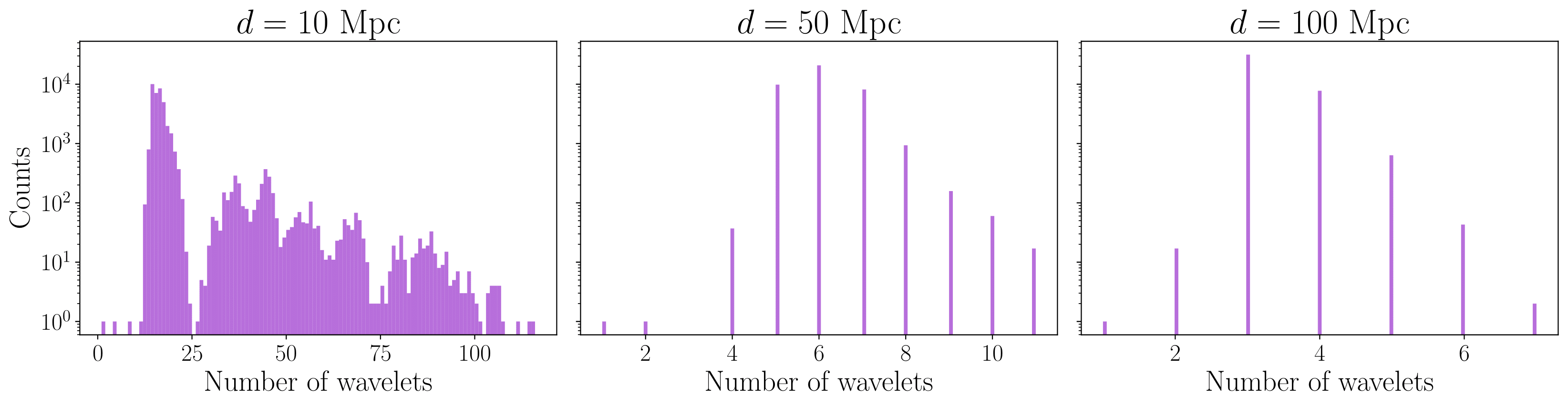}
    \caption{Histograms of the number of wavelets used by \textsc{BayesWave} for the reconstructions of signals injected in ET coming from sources at $d=\{10,50,100\}$ Mpc. The $y-$axis indicates the number of iterations of the RJMCMC algorithm that use a certain number of wavelets. At each iteration, the algorithm might add wavelets to the series, but only when the fit is improved considerably so as to overcome the Occam penalty.}
    \label{fig:wavelets}
\end{figure*}

The complete GW strains and the corresponding amplitude spectral density (ASD) of both injected (red) and recovered (blue) colored time-domain signals for the APR4 EOS model of Table~\ref{tab:table_sim} are depicted in Figs.~\ref{fig:ETTimeWindow} and~\ref{fig:H1TimeWindow}, using the power spectral density (PSD) of ET and H1, respectively, and for a source at a distance of 3~Mpc. The blue-shaded regions show the 50\% and 90\% credible intervals (CIs) of the posterior distribution of the reconstructed signal. The limits of these intervals correspond to the values of the percentiles 25th/75th and 5th/95th, respectively. Time windows with different widths located at different stages of the post-merger phase are applied to the time series. Those are indicated by the areas depicted in yellow in the top panels of both figures. By moving those windows over time we can follow potential changes in the ASD during the evolution of the GW signal, and observe the emergence of different modes in the HMNS. The black dashed line in the bottom row of the two figures corresponds to the ASD of the injected entire signal, from $t_i = -20$ ms (where $t=0$ ms corresponds to the time of merger) to $t_f = 140 $ ms. Correspondingly, the red lines in the ASD plots show the corresponding spectrum for the selected time-window intervals. One can clearly see that the peak frequency changes depending on the time window applied to obtain the ASD, shifting to lower frequencies for increasingly later times. We do not include the corresponding plots for the SLy EOS model because a similar behaviour is observed in this case.

By comparing the two figures the differences between the reconstructions of the injections into H1 and E3 are evident. The early post-merger signal corresponding to the $f$-mode is well recovered for both types of detectors. We note that this is in agreement with the previous findings of~\cite{Chatz:2017} who used BNS merger waveforms from the numerical-relativity  simulations of~\cite{Bauswein:2014,Bauswein:2016} (extending only up to $\sim 15$ ms after merger) to recover with \textsc{BayesWave} the peak frequency  of the $f$-mode. However, when it comes to the late post-merger signal during which the inertial modes are excited, only a third-generation detector such as ET is able to reasonably reconstruct the waveform. We also performed a similar study with Cosmic Explorer (CE) \citep{Evans:2021} finding comparable results. 

The waveform posterior distribution can be used to derive some physical parameters of the HMNS. In this case, the reconstructed signals can be used to obtain the posterior for the dominant post-merger frequency $f_{\rm peak}$~\cite{Bauswein:2012b,Chatz:2017,Bose:2018}. For both the overlap and the reconstructed peak frequency we study both the entire post-merger signal, which is dominated by the $f$-mode excited at early times, and the late signal, during which the inertial modes are excited. For completeness, Appendix A discusses a test case using injections that only contain the late post-merger phase.

\subsubsection{Study of the full GW signal}

We compute the overlap between the injected and the recovered waveforms to test the performance of \textsc{BayesWave}. In Fig.~\ref{fig:overlap_full} we show the overlap as a function of distance for both APR4 and SLy EOS, computed for the HLV detector network (left panel) and for ET (right panel). The overlap clearly decreases with the distance to the source, as the GW signal becomes more difficult to reconstruct. The behaviour is the same for both EOS, but the reconstruction is slightly better for the APR4 EOS at larger distances. Note also the difference between the detector networks: the HLV network has $\mathcal{O}_{\rm network}\sim 0.7$ at $15$~Mpc and the ET gives a similar overlap at roughly $150$~Mpc.
For the sake of comparison, in \cite{Chatz:2017} an overlap of $\sim 0.9$ is reported for a post-merger SNR of 5. In our case, $\mathcal{O}_{\rm network} =0.9$ is achieved for a distance of $\sim 12$ Mpc, which translates to a post-merger SNR of $\sim 5$ for both EOS.

We show the dependence of the recovered value of $f_{\rm peak}$ with the distance to the source in Fig.~\ref{fig:fpeak_full}, again considering the full waveform, $t\in [-20,140]$ ms, injected in a window of 1 second of detector data, which corresponds to colored Gaussian noise. This value,  plotted with solid curves, is the mean value obtained from the posterior distributions of the recovered signals. We also depict the standard deviations for both EOS, which become larger as the distance to the source increases. These results are consistent with the overlap values shown in Fig.~\ref{fig:overlap_full}, since a low overlap value gives a poorly recovered $f_{\rm peak}$. In the case of H1 (left panel), the dispersion starts increasing at $d\sim 13$ Mpc for APR4 and $d\sim 17$ Mpc for SLy, right where $\mathcal{O}_{\rm network}$ drops below 0.6. Concerning ET (right panel), the uncertainty becomes larger at $d\sim 125$ Mpc, also when $\mathcal{O}_{\rm network} \sim 0.6$. Notice that for high SNR the distance to the source and the SNR are inversely proportional, and a less accurate value of $f_{\rm peak}$ would be obtained by decreasing the SNR.

In~\cite{Chatz:2017} an almost flat posterior distribution for a post-merger SNR of 3 was obtained.
In our case, at 25 Mpc the recovery of the frequency peak already has a large uncertainty, and corresponds to a post-merger SNR of $\sim 3$ for both EOS, which is consistent with the results of~\cite{Chatz:2017}.

In Fig.~\ref{fig:wavelets} we depict histograms of the numbers of wavelets used for the reconstructions at different distances, $d=\{10,50,100\}$ Mpc. The closer the source the larger number of wavelets are employed, resulting in more accurate reconstructions.

\subsubsection{Study of the late post-merger phase}

\begin{figure*}[ht]
\centering
   \includegraphics[width=0.32\textwidth]{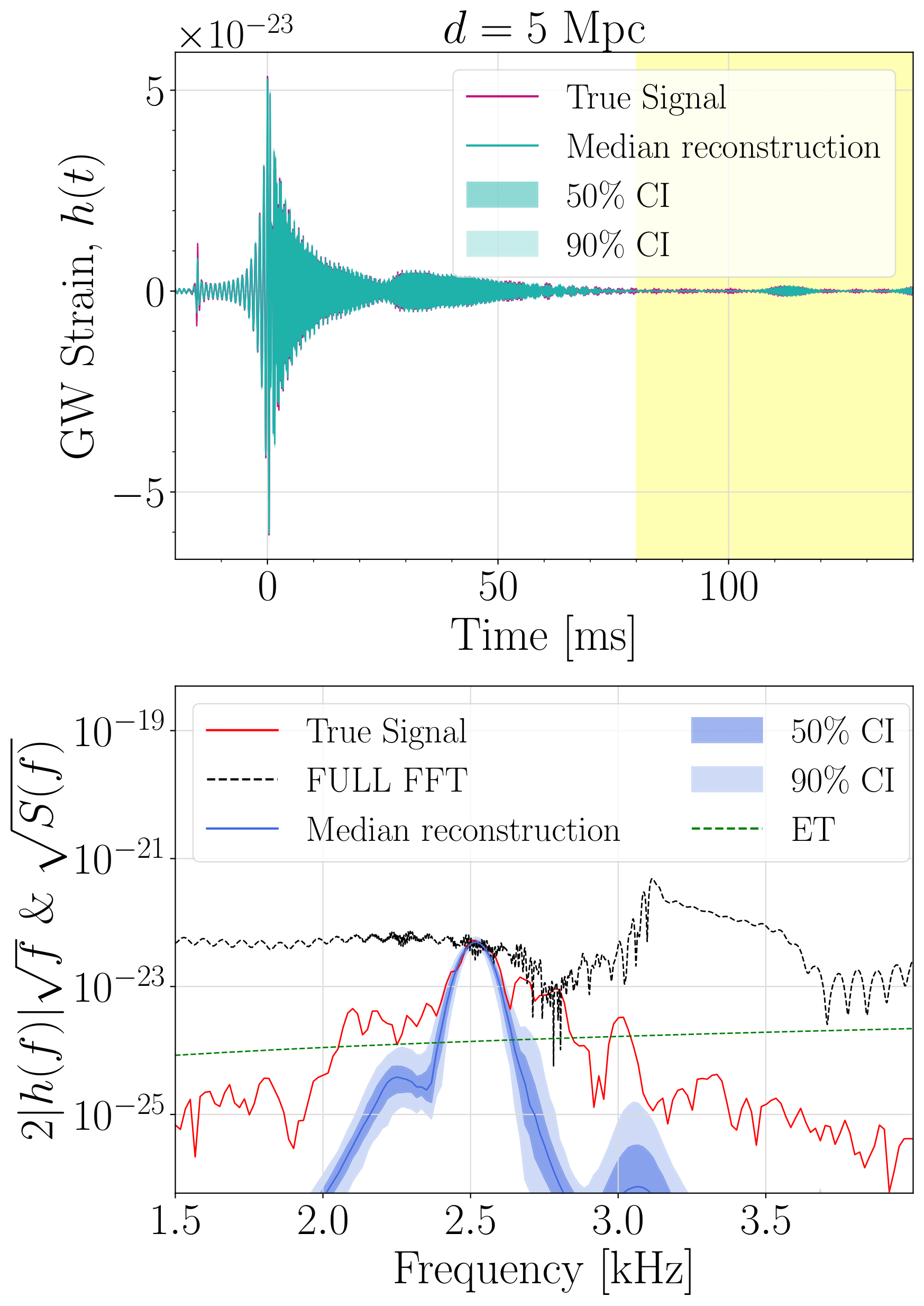}
   \includegraphics[width=0.32\textwidth]{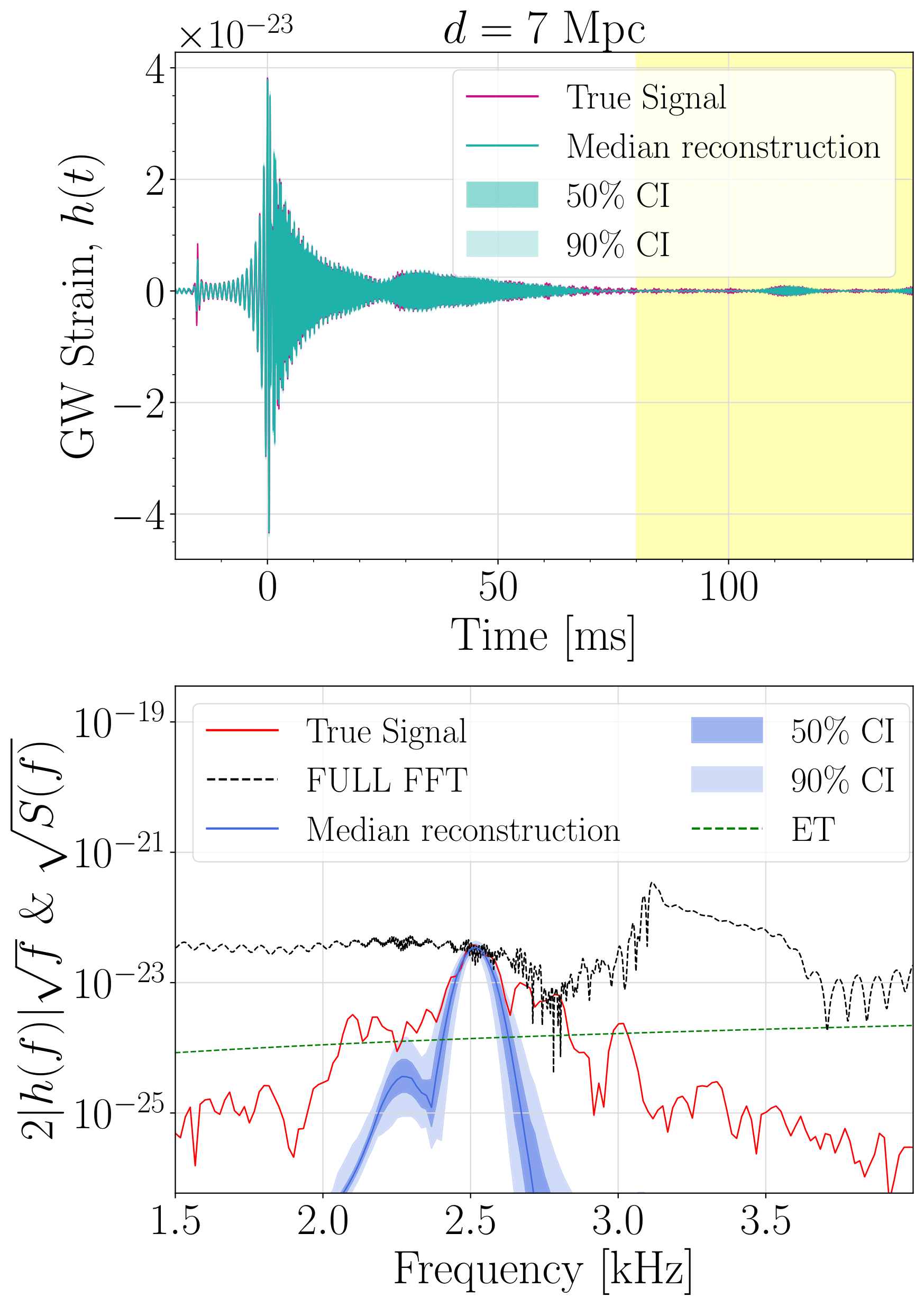}
   \includegraphics[width=0.32\textwidth]{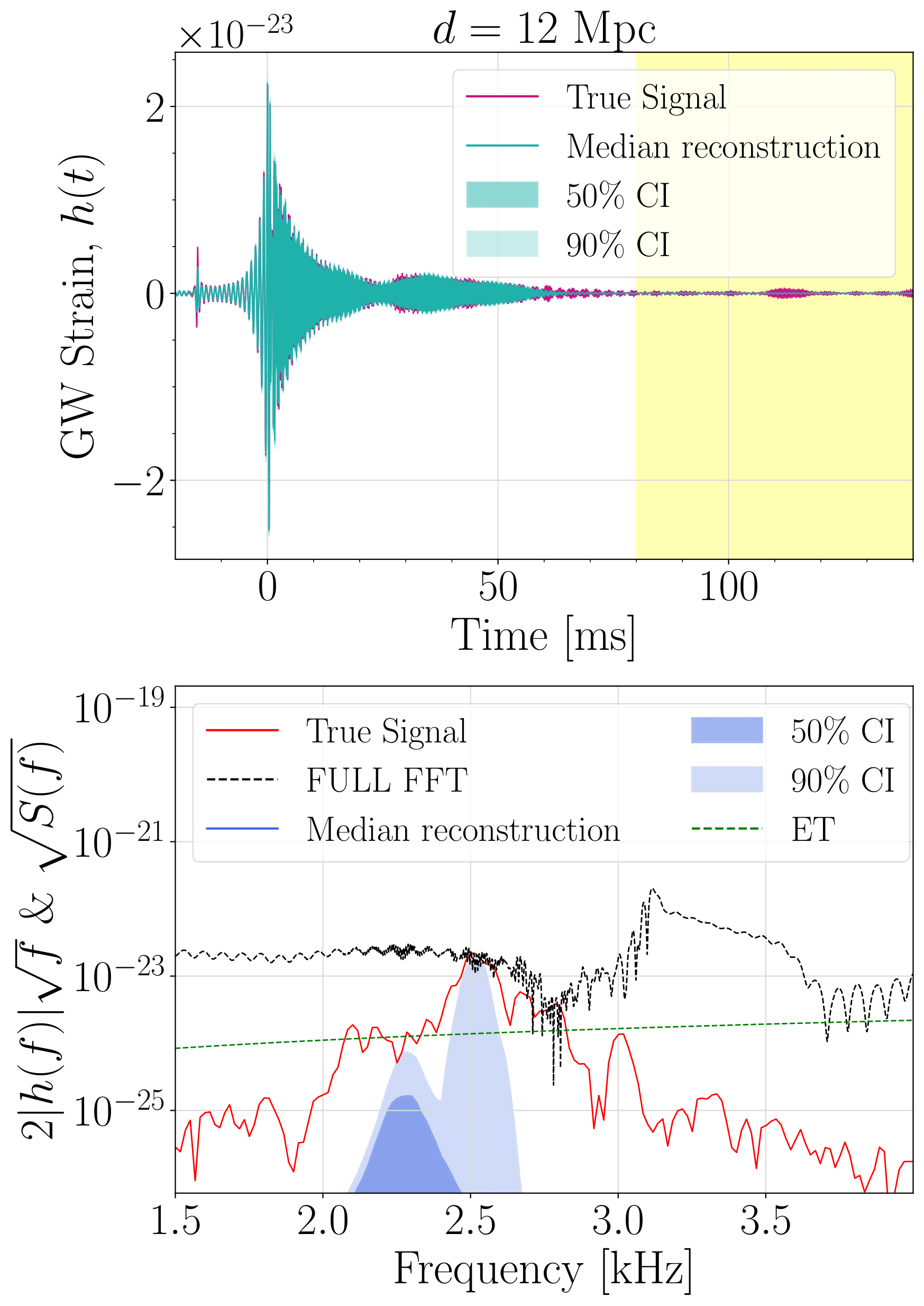}
\caption{Injected (red) and recovered (blue) time-domain waveforms (top panels) and ASD (bottom panels) for sources located at $d=\{5,7,12\}$ Mpc. Signals are injected in ET (E3). The time window used to compute the ASD only considers the last part of the signal (yellow region). The recovery of the frequency peak degrades with the distance to the source.}
\label{fig:rec_late_ET}
\end{figure*}

\begin{figure*}[ht]
    \centering
   \includegraphics[width=0.32\textwidth]{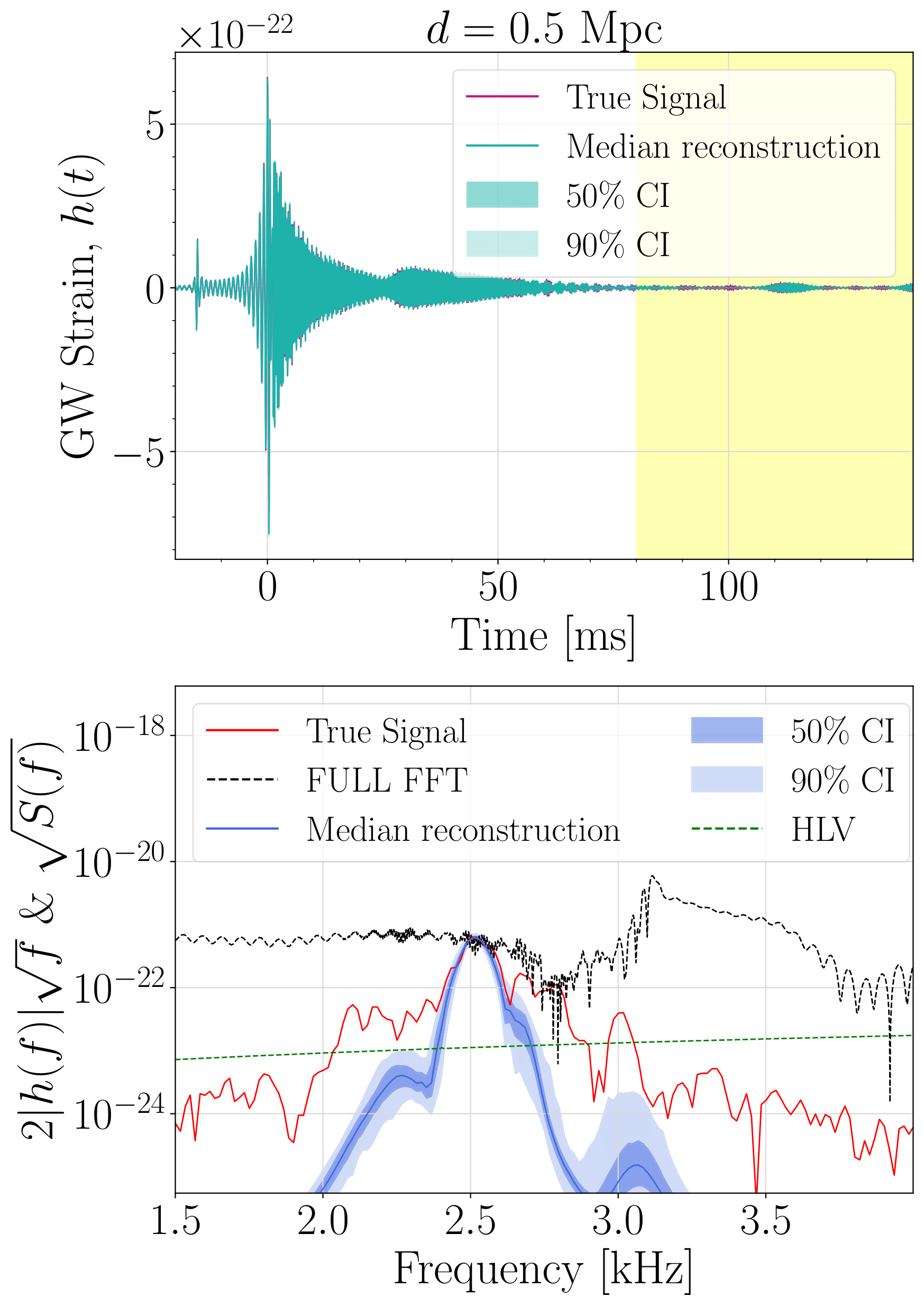}
     \includegraphics[width=0.32\textwidth]{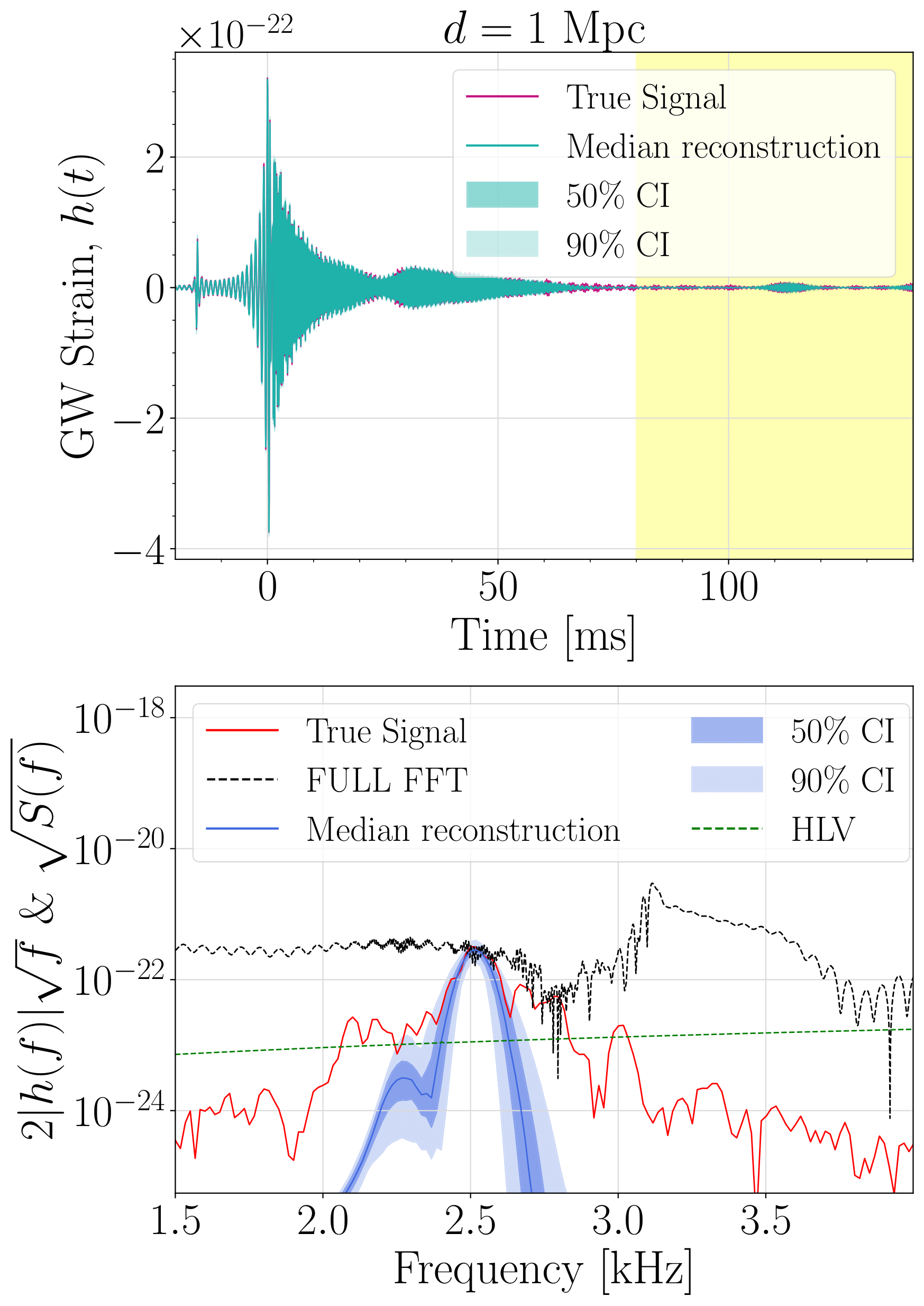}
   \includegraphics[width=0.32\textwidth]{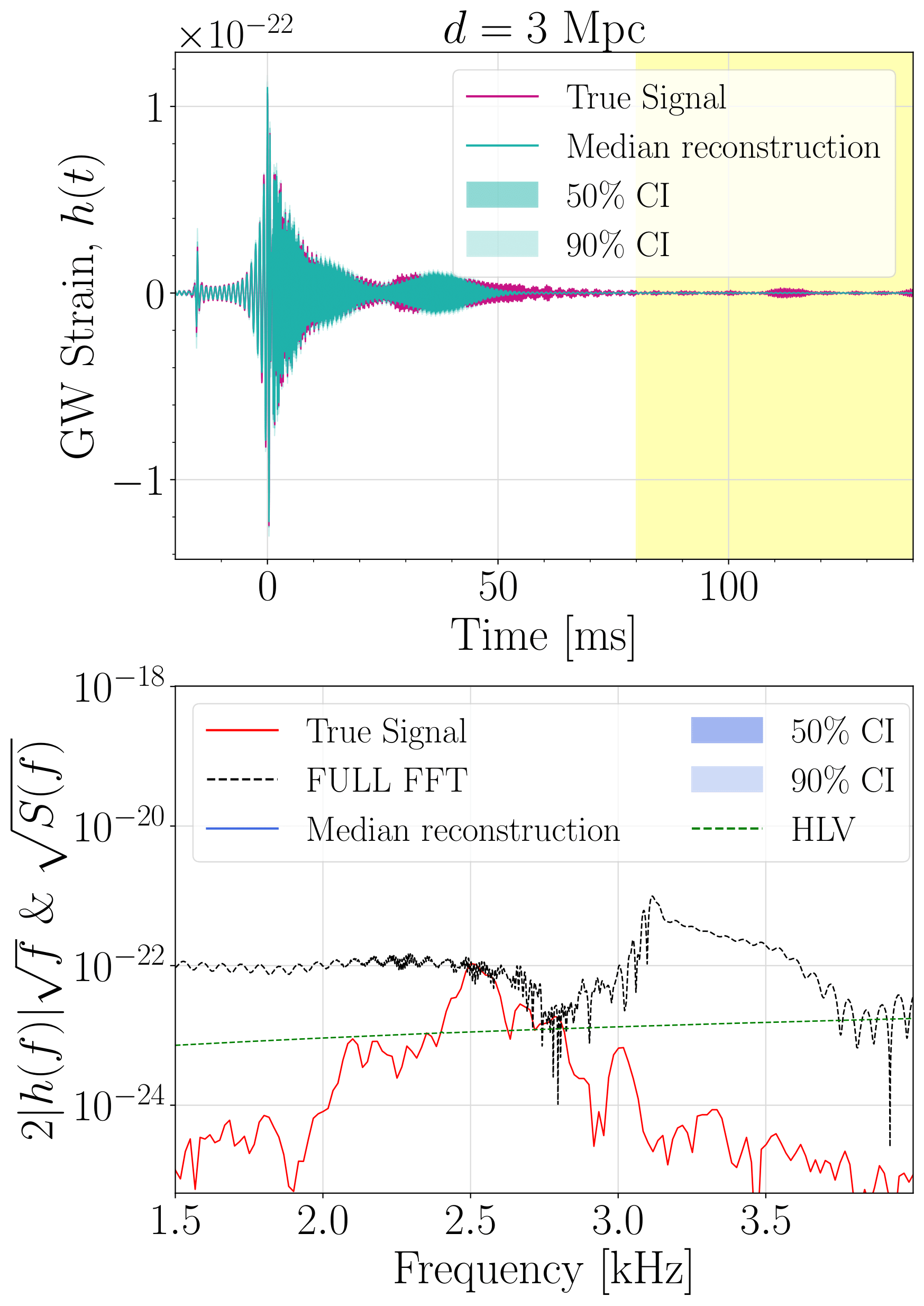}
\caption{As Fig.~\ref{fig:rec_late_ET} but for H1 and closer source distances, $d=\{0.5,1,3\}$ Mpc}.
\label{fig:rec_late_HLV}
\end{figure*}

\begin{figure*}[ht]
\centering
   \includegraphics[width=0.95\linewidth]{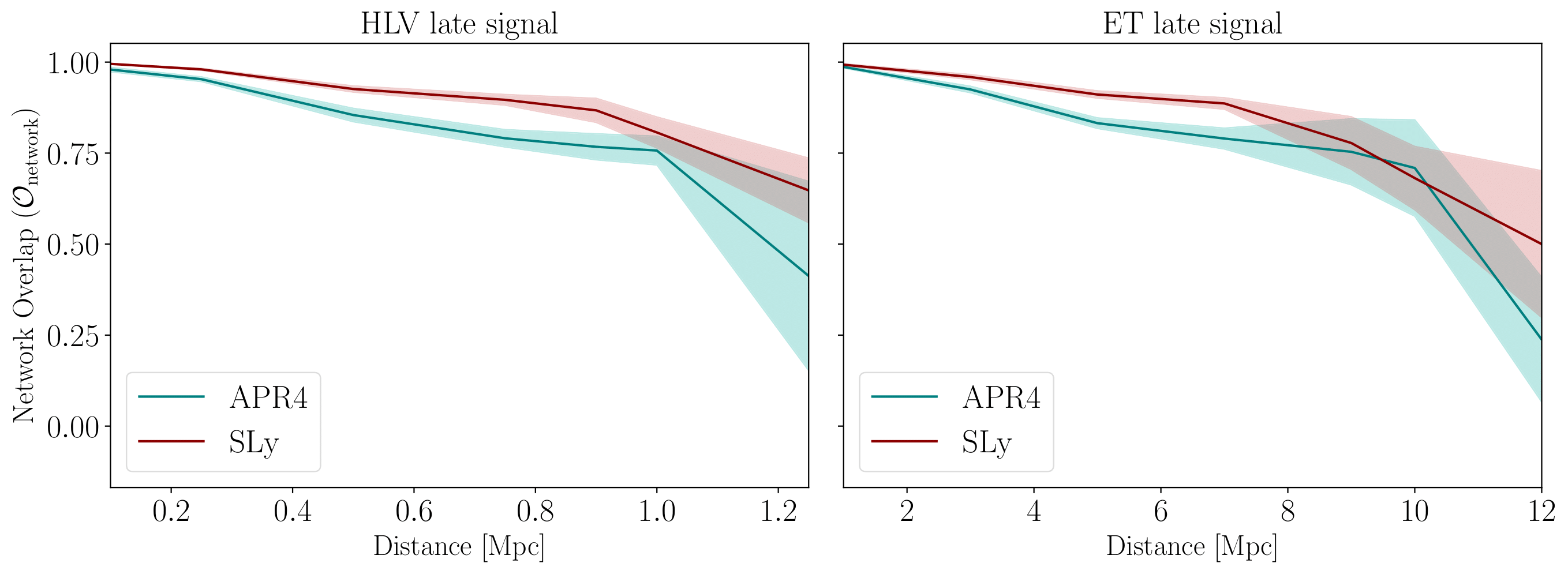}
\caption{Detector overlap of the late post-merger GW signals for the HLV network (left panel) and ET (right panel). Solid lines represent the mean values from the posterior distributions and shaded regions are the standard deviations. For both EOS the overlaps drop below $\approx 0.75$ for a source distance of about 1 Mpc for HLV and 10 Mpc for ET.}
\label{fig:overlap_late}
\end{figure*}

\begin{figure*}[ht]
\centering
   \includegraphics[width=0.95\linewidth]{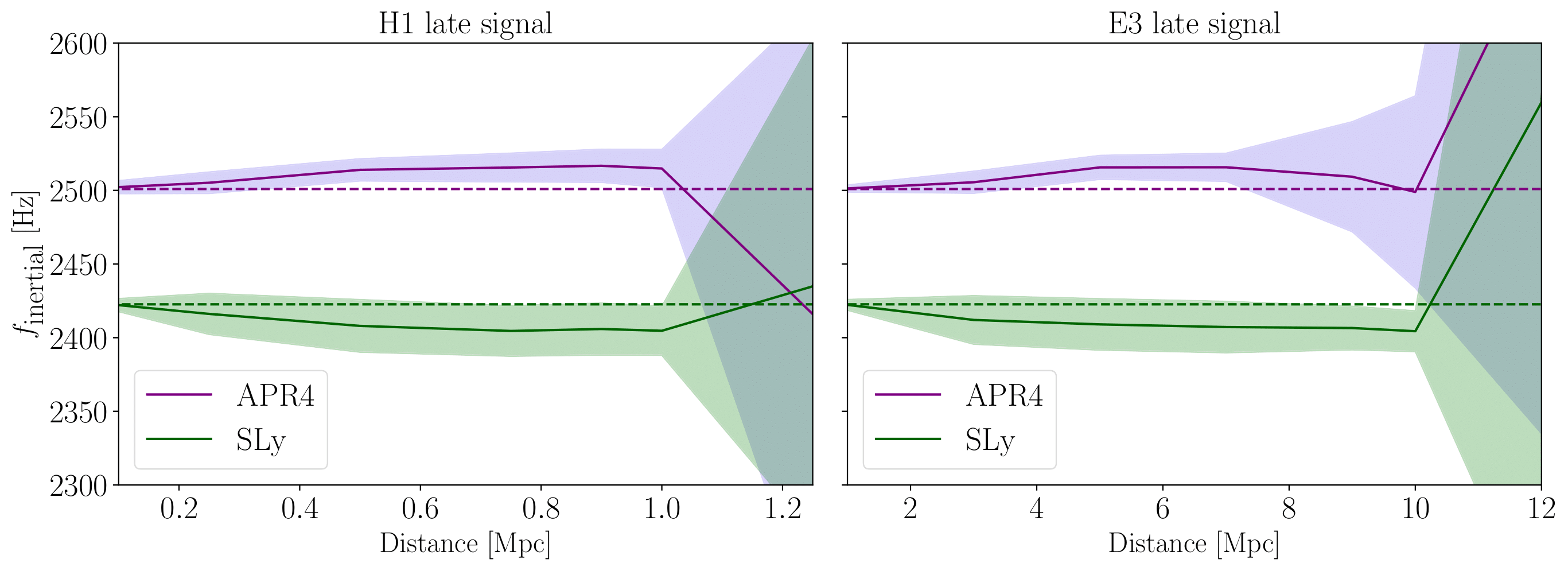}
\caption{Dependence with distance of the peak frequency during the late post-merger phase, for signals injected in H1 (left panel) and ET (right panel). Solid lines and shaded areas are the mean values and the standard deviations of the distributions, respectively. For ET, the peak frequency is well identified up to 10~Mpc for both APR4 and SLy EOS  while for H1 a satisfactory recovery is only possible for sources up to 1~Mpc for both EOS.}
\label{fig:fpeak_late}
\end{figure*}

We turn next to analyze the reconstruction of the late post-merger signal ($t\geq 40$ ms for SLy EOS and $t\geq 80$ ms for APR4 EOS \cite{DePietri:2020}). Once the maximum amplitude of the fundamental quadrupolar $f$-mode has significantly decreased, a signal with lower frequency and amplitude appears, associated with the manifestation of inertial modes in the remnant. The two rightmost panels of Fig.~\ref{fig:ETTimeWindow} clearly show the appearance of this new peak for a source observed at a distance of 3 Mpc. This peak is just above the sensitivity curve of ET at a frequency of $\sim 2500$ Hz (see rightmost panel of Fig.~\ref{fig:H1TimeWindow}). However, it is out of reach for current detectors, at least for $d= 3$ Mpc (cf.~Fig.~\ref{fig:H1TimeWindow}). 

To illustrate how the \textsc{BayesWave} reconstruction changes with distance, we show in Fig.~\ref{fig:rec_late_ET} the injected and reconstructed time-domain waveforms and the respective ASD reconstructions for three representative distances, namely 5, 7, and 12 Mpc, and for  BNS merger simulations with the APR4 EOS. As before, the regions in yellow in the top panels show the time window we use to compute the ASD displayed in the bottom panels. The median of the reconstructed ASD is shown with a blue solid line and the 50\% and 90\% credible intervals are indicated by the dark and light blue-shaded areas, respectively. The Hann function (as other window functions) cuts the tails of the time-domain signal, and thus might affect the resulting frequency spectra. However, the inertial modes with largest amplitude are located in the middle of the time window, and are not affected by the cut. We find that the region around the frequency peak at $f \approx 2.5$ kHz is well recovered when the source is at 5 Mpc. On the other hand, as the distance increases the reconstruction worsens, as expected, and for a source at 12 Mpc there is no frequency peak in the reconstructed signal. The corresponding result for current detectors is shown in Fig.~\ref{fig:rec_late_HLV} which depicts the dependence of the peak-frequency recovery with distance (for $d=\{0.5, 1,3\}$ Mpc) with the design sensitivity of H1. In this case, \textsc{BayesWave} is not able to recover the peak-frequency of inertial modes even when the GW source is at 3~Mpc.

In Fig.~\ref{fig:overlap_late} we show the network overlap function of the late post-merger signal, for both HLV and ET. The same initial time windows as in Figs.~\ref{fig:rec_late_ET} and \ref{fig:rec_late_HLV} are used to compute these overlaps. We note, however, that the final time of the window is different for both EOS, namely 140 ms for the APR4 EOS and 123 ms for the SLy EOS, respectively. For the latter the final time is shorter since a black hole forms at $t\approx 123.6$ ms~\citep{DePietri:2020}. Moreover, the initial time of the window is also different, as pointed out before, since the emergence of the inertial modes occurs at different times ($t\sim 40$ ms for SLy and $t\sim 80$ ms for APR4). For the case of ET (right panel), at $d \approx 8$ Mpc the overlap is around 0.7 for both EOS, but it rapidly decreases to about 0.5 at $\approx 12.5$ Mpc. The SLy EOS gives a higher overlap and a more accurate $f_{\rm inertial}$ but both EOS yield $\mathcal{O}_{\rm network} =0$ at 15 Mpc. On the other hand, for the HLV detector network the network overlap for both EOS falls rapidly to practically 0 from a distance of 1.75 Mpc.

We now focus on the recovery of the frequency peak of these lower-frequency inertial modes, $f_{\rm inertial}$. Fig.~\ref{fig:fpeak_late} depicts the dependence of the recovered $f_{\rm inertial}$ with distance for H1 and ET. The maximum distance shown in the plots for each detector is selected by the value at which the reconstructions start to significantly fail. 
These results are in good agreement with the overlap shown in Fig.~\ref{fig:overlap_late}. We note that there is a slight dependence on the EOS as we obtain a peak frequency that is about 75 Hz higher in the APR4 case. For the specific case of ET, at $d=12$~Mpc the recovery of the peak frequency fails for both EOS. Up to 8 Mpc, the recovered $f_{\rm inertial}$ is close to the injected one with an uncertainty of $\Delta f_{\rm inertial} \lesssim 25$ Hz. For the case of H1, this value of the uncertainty of the method is obtained for much shorter distances ($d\approx 1.0$~Mpc).

\section{Conclusions}
\label{sec:discussion}

The existence of convectively unstable regions in long-lived remnants of BNS mergers~\citep{DePietri:2018,DePietri:2020} triggers the excitation of inertial modes, which depend on the rotational and thermal properties of the remnant. Their presence in the late post-merger GW signal might thus provide further insight in our understanding of neutron star properties. In this paper we have studied the possibility of reconstructing the late BNS post-merger GW signal with current and future interferometers. To this aim we have employed the waveforms produced in numerical-relativity simulations of equal-mass BNS mergers that last up to $t\approx 140$ ms after merger, performed by \cite{DePietri:2018,DePietri:2020}. These long-lasting simulations showed the excitation of oscillation modes in the post-merger remnant with a smaller frequency and amplitude than those of the quadrupolar $f$-mode which dominates the GW spectra of the early post-merger phase. These so-called inertial modes are triggered by a convective instability developing in the HMNS, for which the Coriolis force acts as the dominant restoring force \cite{Stergioulas:2004,Kastaun:2008,Stergioulas:2011}. The late-time appearance of these modes has also been observed in the BNS simulations of~\cite{Ciolfi:2019} accounting for the effects of magnetic fields in the dynamics.

Due to their small amplitude, with a strain $h(f)$ more than one order of magnitude smaller than that of the $f$-mode, the detectability of such inertial modes can be challenging. In order to assess their possible detection, we have employed the $\textsc{BayesWave}$ algorithm \cite{Cornish:2015,Littenberg:2015} to reconstruct our time-domain waveforms injected into Gaussian noise. The signals were injected at different distances from the source to check the range of detection of those modes. In all cases the source was assumed to be optimally oriented with respect to (one of) the detectors. 

Our study reveals that current GW interferometers (i.e.~the HLV network) are able to recover the peak frequency of inertial modes only if the BNS merger occurs at distances of about 1 Mpc or less. However, for future detectors such as ET, the range of detection increases by a factor of 10, consistent with their increased sensitivity compared to current detectors. An important point to stress is that the difference between the frequency peaks of the inertial modes for different EOS (APR4 and SLy) is bigger than the difference between the peaks of the fundamental mode in the early part of the signal. This means that a future detection of those late post-merger modes could give us more insight into the internal matter {and structure} of a neutron star, as a result of the broken EOS degeneracy and the relationship of those modes with the rotational properties of differentially rotating stars. 
In general the frequency $f_{\rm inertial}$ changes with the EOS and the total binary mass and it also  correlates with the tidal deformability. For the simulations discussed in this work $f_{\rm inertial}$ appears to be very close for all models because of the properties of the initial systems, in particular the total mass.
Employing different initial data with a wider spread in the total mass might be something
worth trying in a future investigation. Furthermore, the value of the peak frequency can be used to infer different physical parameters of the star \cite{Kastaun:2008}, extending what has already been done for the $f$-mode to infer the radius, the tidal coupling constant or the average density of the neutron star \cite{Bauswein:2012,Takami:2015,Bernuzzi:2015,Chatz:2017}. However, as mentioned in \cite{DePietri:2018}, one would need to employ  perturbative studies to identify the particular inertial modes that are excited. Such a challenging project is outside of the scope of this work, which has purely focused on the prospects of detectability of inertial modes.

\begin{acknowledgments}
We thank the anonymous referee for useful remarks.
We also thank Sudarshan Ghonge for useful discussions and for sharing some \textsc{Python} scripts with us, and Katerina Chatziioannou, James A. Clark, Meg Millhouse, Argyro Sasli, Nikolaos Stergioulas, Juan Calderón Bustillo and Alejandro Torres-Forn\'e for useful comments. The authors are grateful for the computational resources provided by the LIGO Laboratory and supported by the U.S. National Science Foundation Grants PHY-0757058 and PHY-0823459, as well as resources from the Gravitational Wave Open Science Center, a service of the LIGO Laboratory, the LIGO Scientific Collaboration and the Virgo Collaboration. We are grateful for computational resources provided by the Leonard E Parker
Center for Gravitation, Cosmology and Astrophysics at the University of
Wisconsin-Milwaukee. Virgo is funded, through the European Gravitational Observatory (EGO), by the French Centre National de Recherche Scientifique (CNRS), the Italian Istituto Nazionale di Fisica Nucleare (INFN) and the Dutch Nikhef, with contributions by institutions from Belgium, Germany, Greece, Hungary, Ireland, Japan, Monaco, Poland, Portugal, and Spain.
This work has been supported by the Spanish Agencia Estatal de Investigaci\'on (Grants No. PGC2018-095984-B-I00 and PID2021-125485NB-C21) funded by MCIN/AEI/10.13039/501100011033 and ERDF A way of making Europe, by MCIN and Generalitat Valenciana with funding from European Union NextGenerationEU (PRTR-C17.I1, Grant ASFAE/2022/003), by the Generalitat Valenciana (PROMETEO/2019/071), and by the European Union’s Horizon 2020 research and innovation (RISE) programme (H2020-MSCA-RISE-2017 GrantNo.~FunFiCO-777740). MMT acknowledges support by the Ministerio de Universidades del Gobierno de Espa\~na (Spanish Ministry of Universities) through the ``Ayuda para la Formación de Profesorado Universitario" No.~FPU19/01750.

\end{acknowledgments}

\appendix

\begin{figure*}
\centering
   \includegraphics[width=0.32\textwidth]{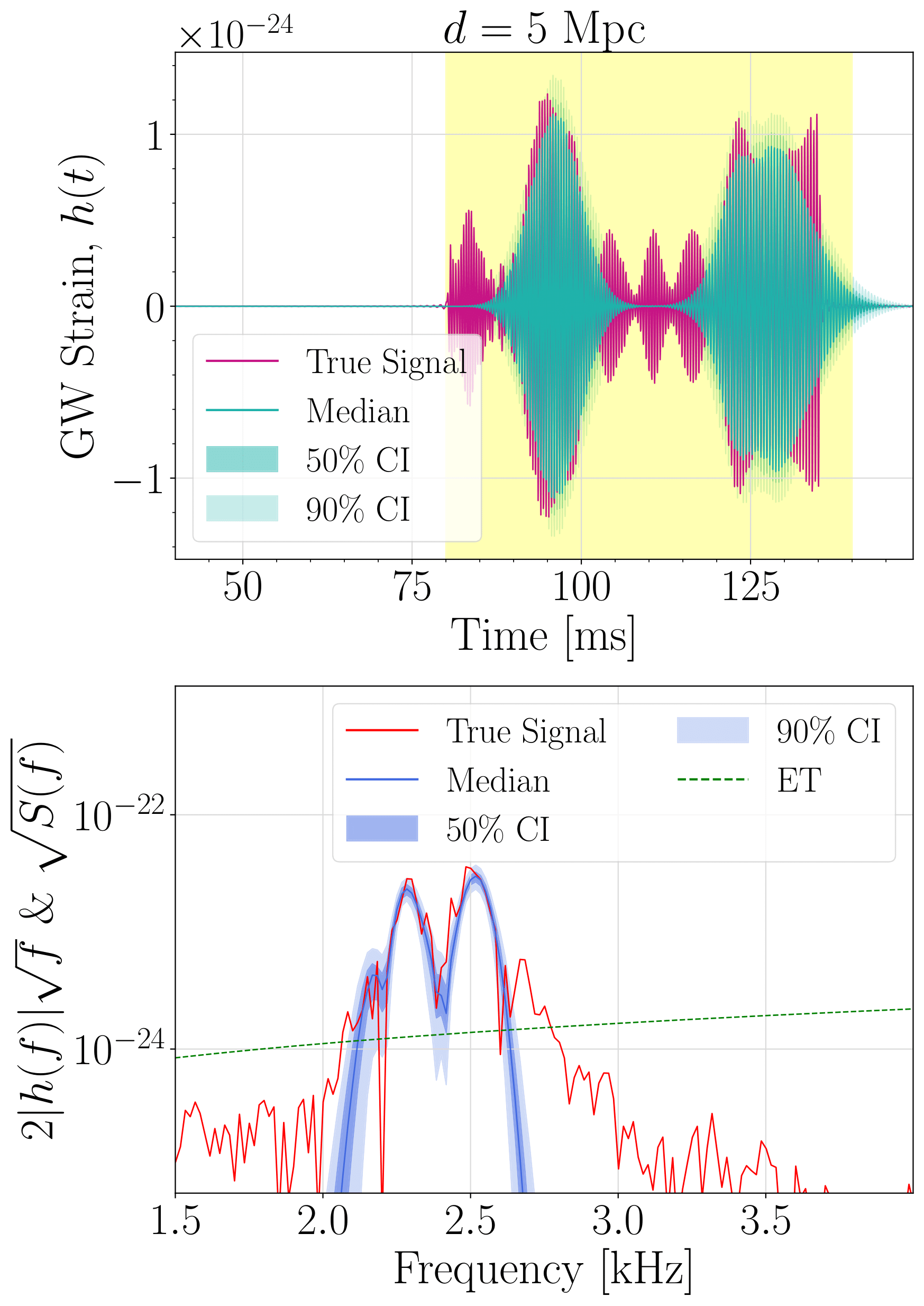}
   \includegraphics[width=0.32\textwidth]{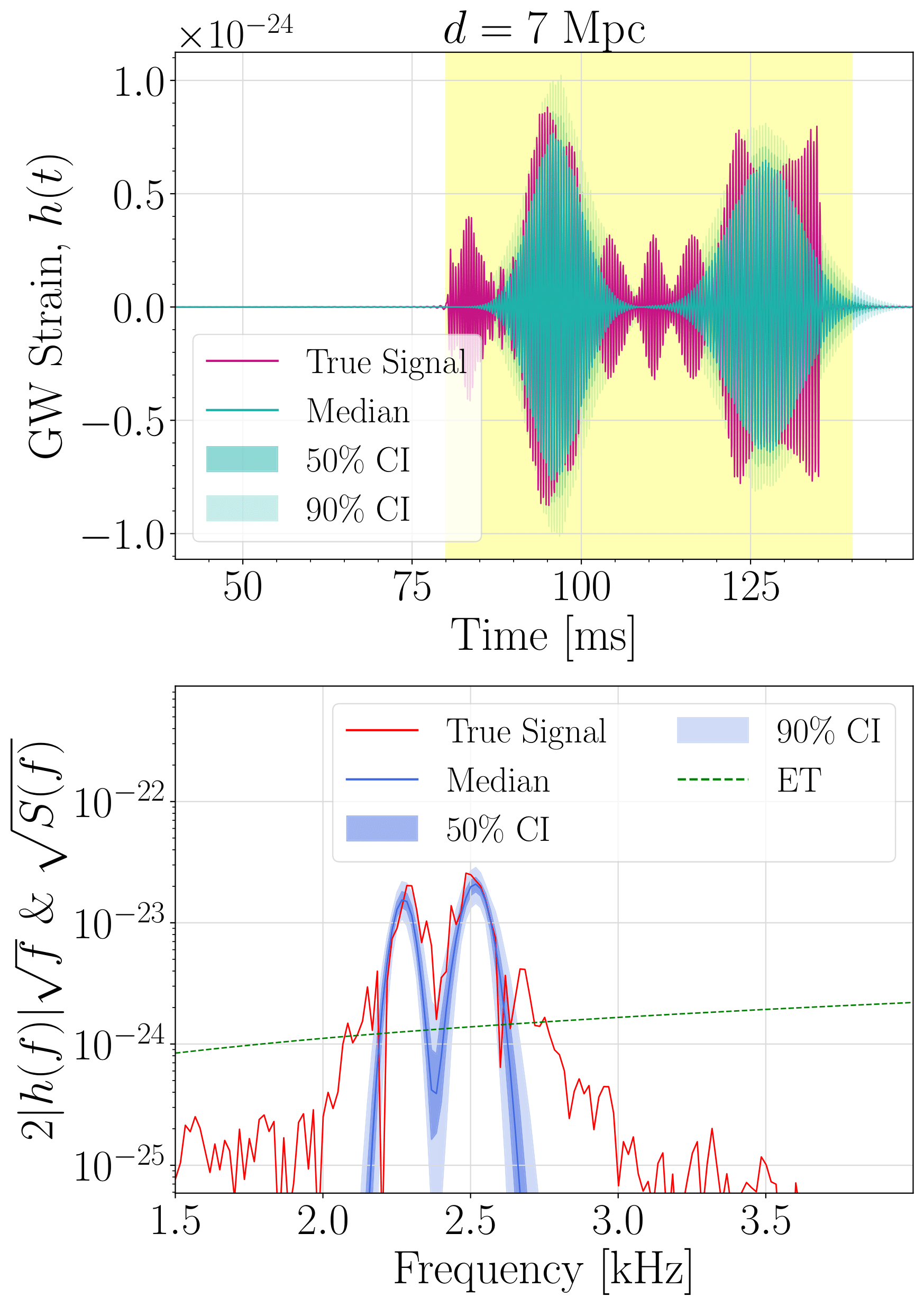}
   \includegraphics[width=0.32\textwidth]{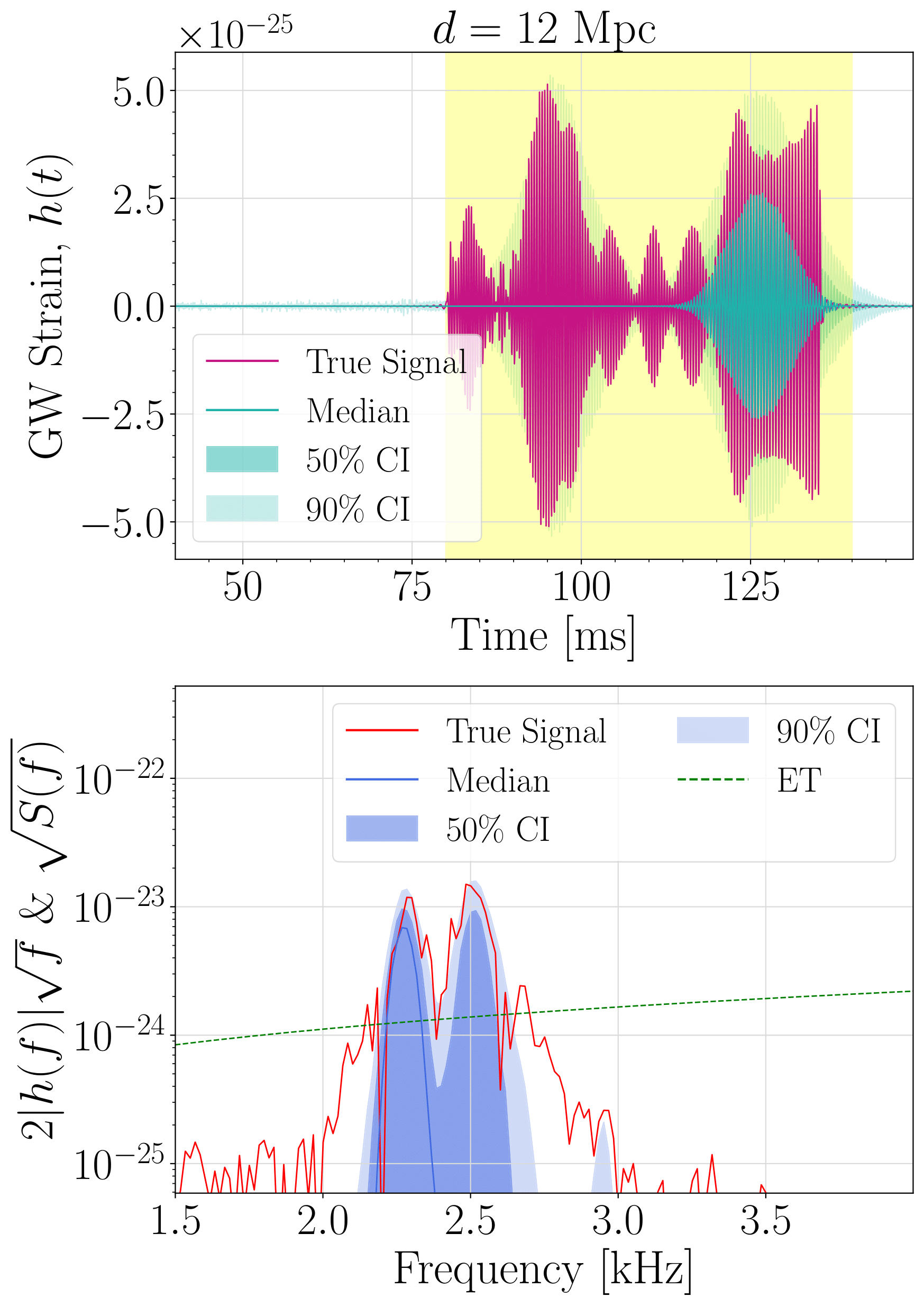}
\caption{Injected (red) and recovered (blue) time-domain waveforms (top panels) and ASD (bottom panels) for sources located at $d=\{5,7,12\}$ Mpc and for APR4 EOS. In this case the injected signal only contains the intertial-mode emission.}
\label{app:appendix_reco_APR4}
\end{figure*}

\begin{figure*}
\centering
   \includegraphics[width=0.32\textwidth]{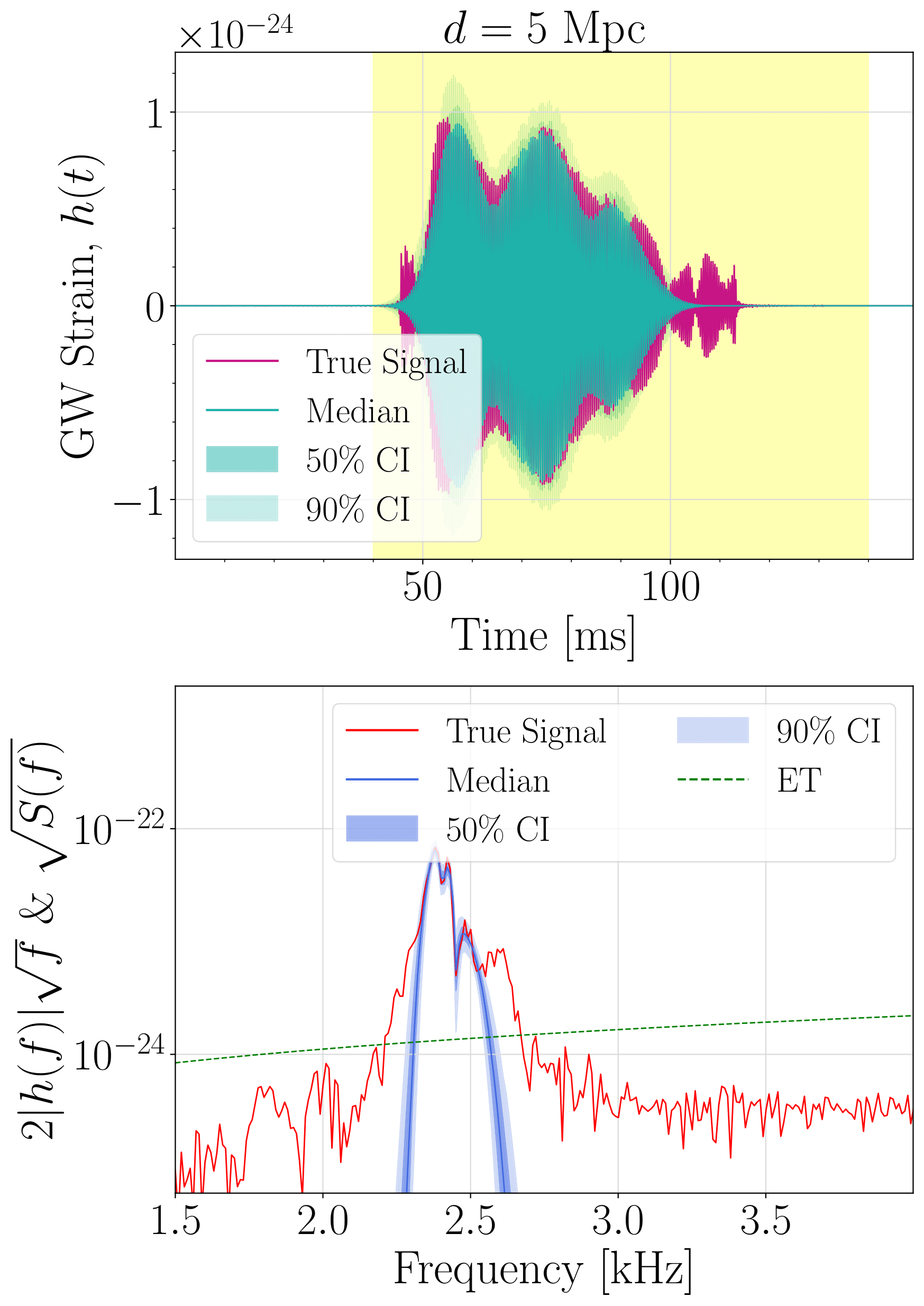}
   \includegraphics[width=0.32\textwidth]{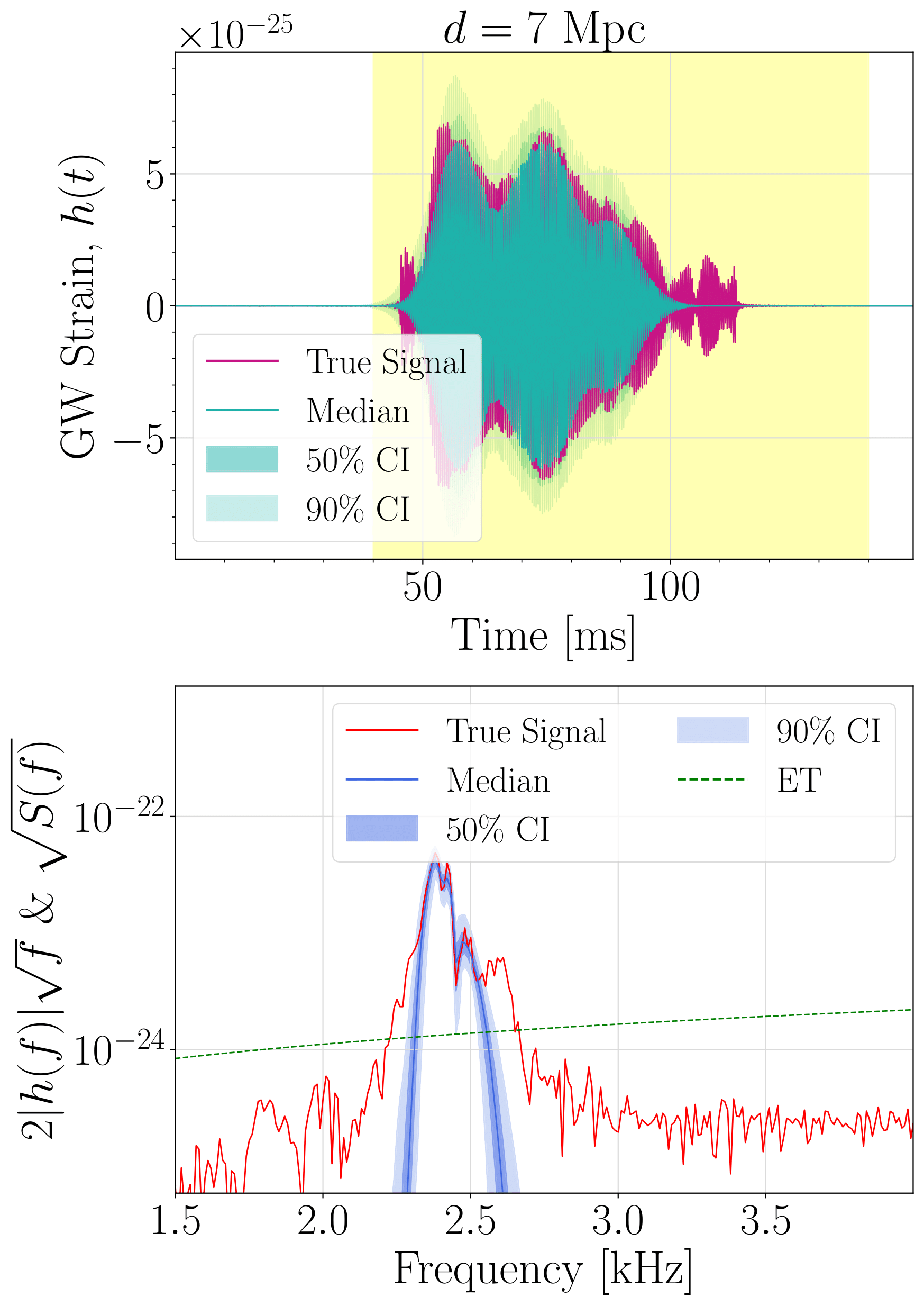}
   \includegraphics[width=0.32\textwidth]{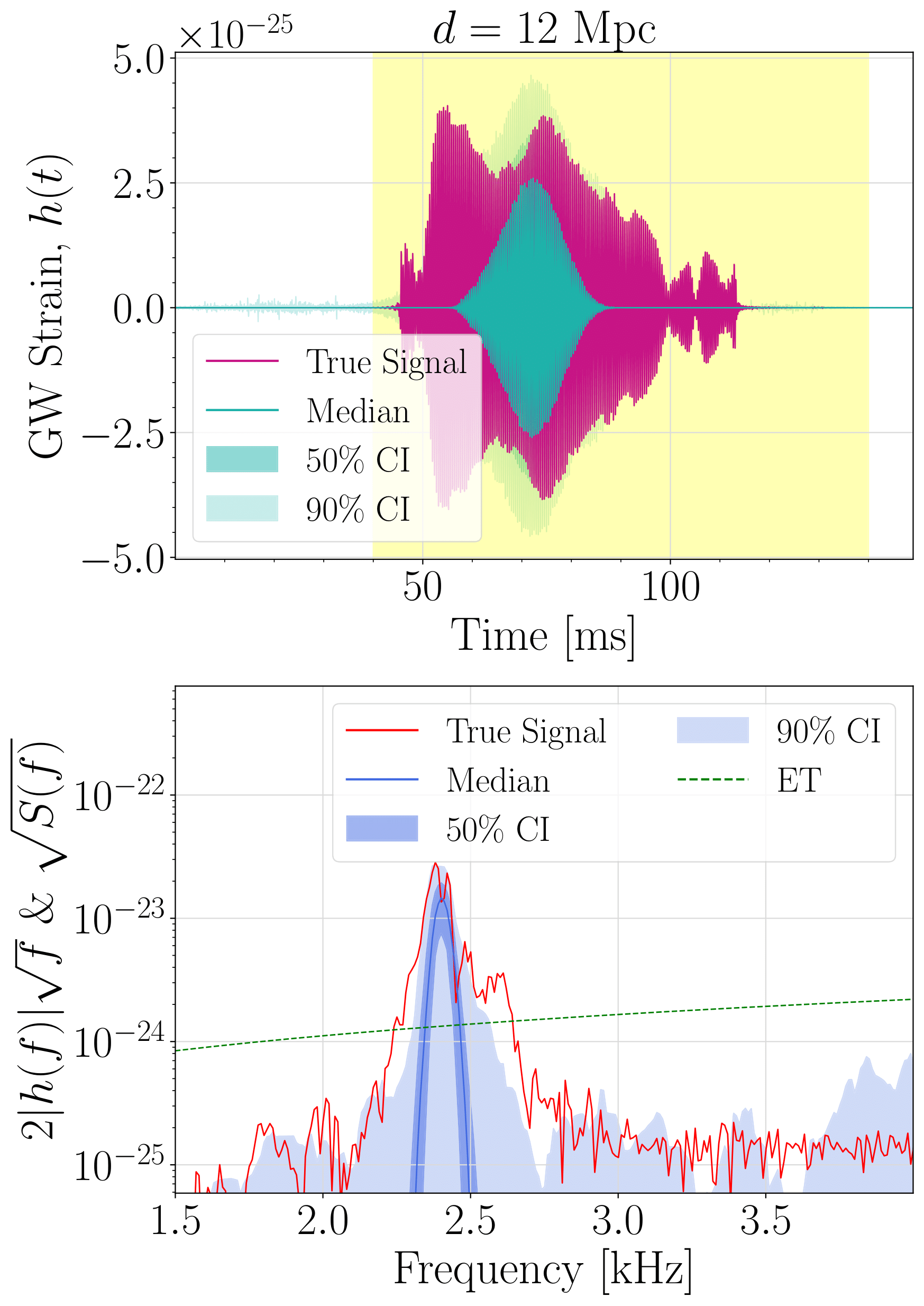}
\caption{As Fig.~\ref{app:appendix_reco_APR4} but for the SLy EOS.}
\label{app:appendix_reco_SLy}
\end{figure*}

\begin{figure*}
    \centering
    \includegraphics[width=0.95\textwidth]{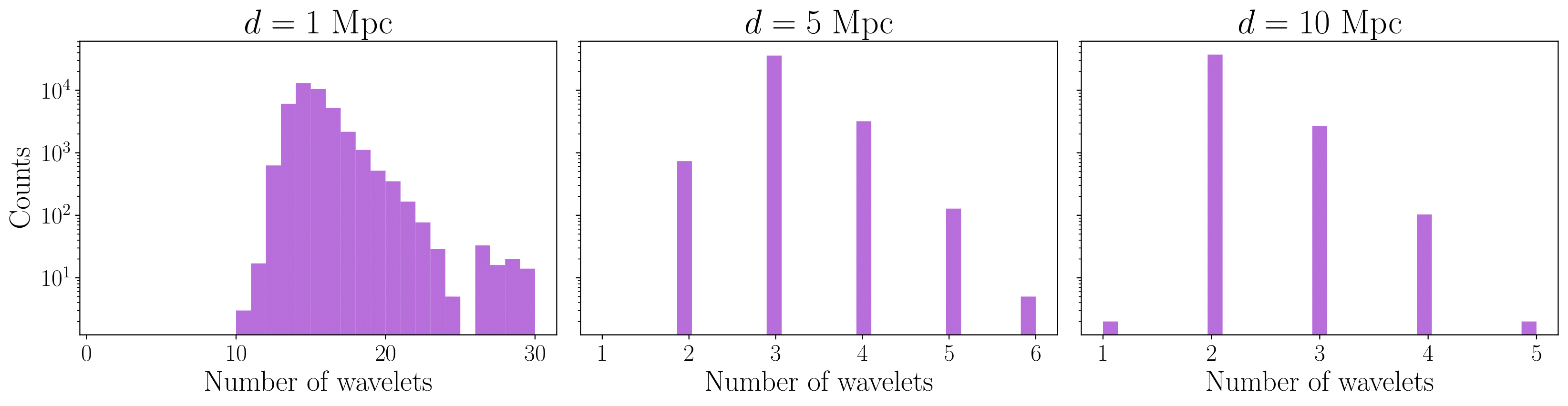}
    \caption{Histograms of the number of wavelets used by \textsc{BayesWave} for the reconstructions of injected signals containing only the intertial-mode emission and located at distances $d=\{1,5,10\}$ Mpc. The $y-$axis indicates the number of iterations that \textsc{BayesWave} uses to build the waveforms by model selection.}
    \label{app:appendix_wavelets}
\end{figure*}

\begin{figure}[t]
\centering
   \includegraphics[width=0.95\linewidth]{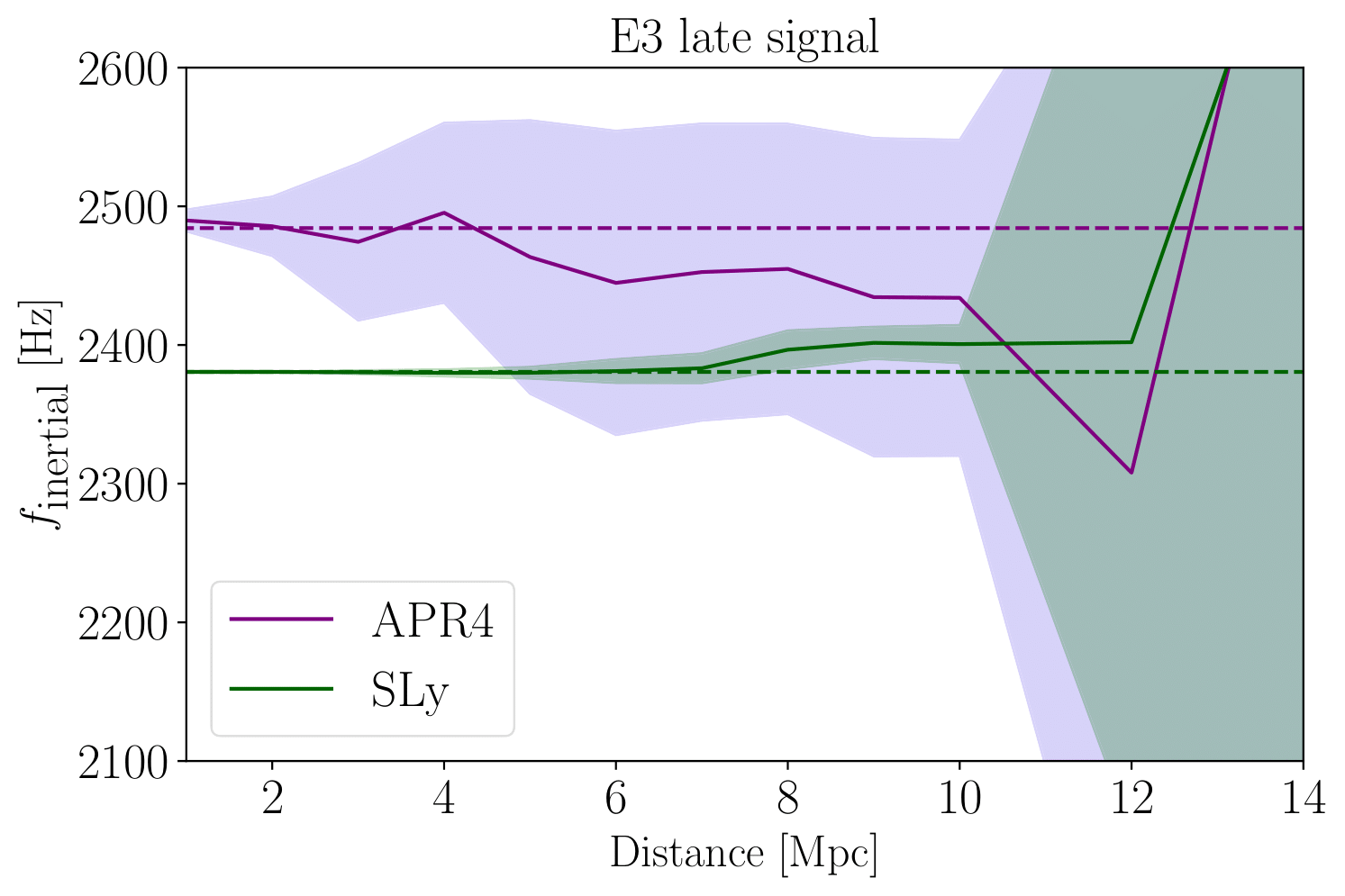}
\caption{Dependence of the recovered frequency peak of the inertial modes with distance for both EOS, when injecting only the part of the post-merger signal corresponding to the inertial modes. Solid lines represent the mean values from the posterior distributions and shaded areas are the standard deviations.}
\label{app:appendix_fpeak}
\end{figure}

\begin{figure}[t!]
\centering
   \includegraphics[width=0.95\linewidth]{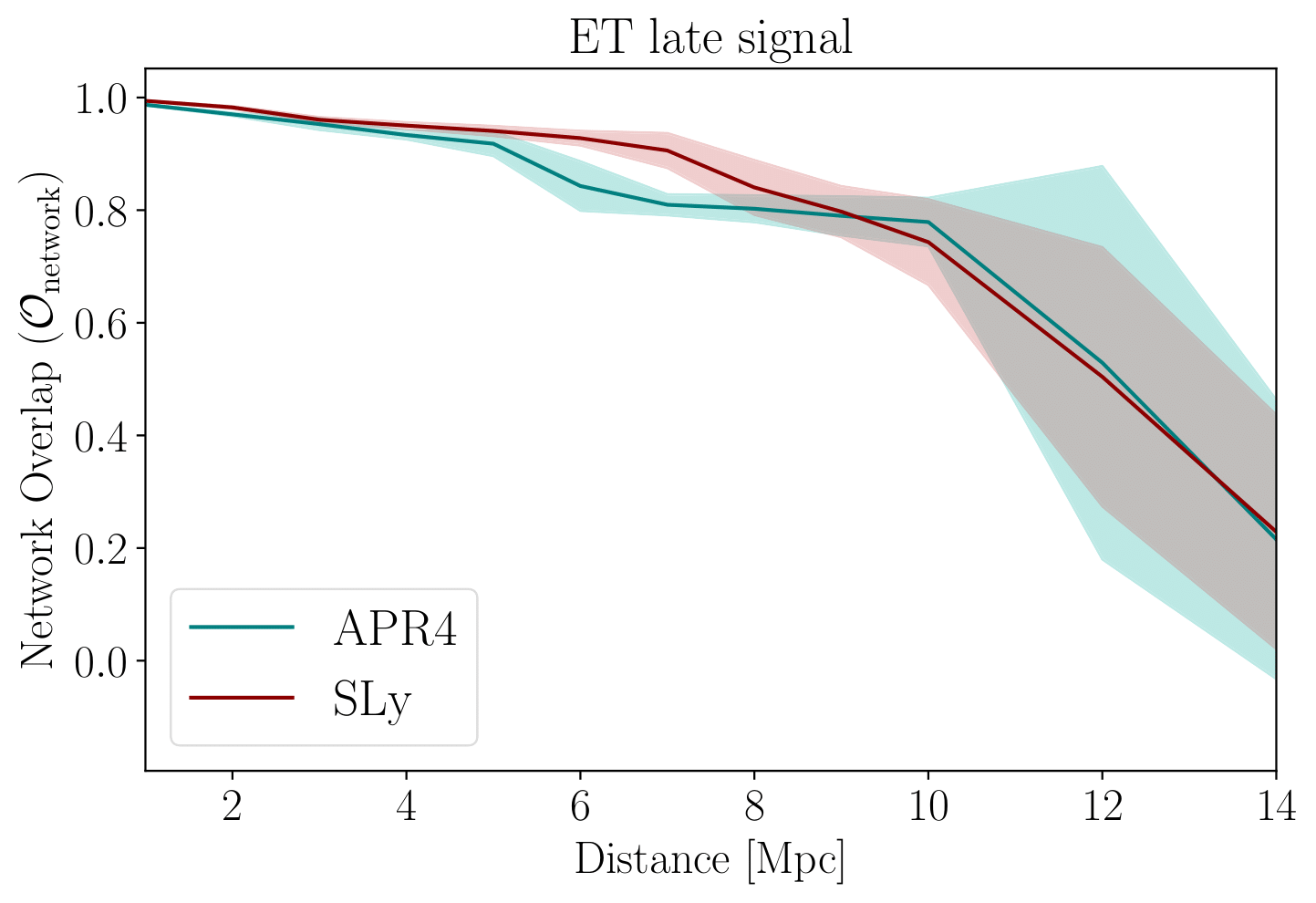}
\caption{Evolution of the overlap between the injected and recovered GW signals with distance for the two EOS. Solid line represent the mean values from the posterior distributions and shaded areas are the standard deviations.}
\label{app:appendix_overlap}
\end{figure}

\section{Reconstruction of late post-merger injections}

In this Appendix we consider injections that only contain the late post-merger phase when inertial modes are active. This test case allows us to assess the capability of \textsc{BayesWave} of recovering only the part of the GW signal containing the inertial-mode emission and to find out whether there is an improvement with respect to the case of full-signal injections discussed in the main text. For APR4 we inject the signal from 80 ms to 140 ms after merger while for SLy the respective range goes from 45 ms to 140 ms after merger. For this test we only consider the ET detector. 

In Figs.~\ref{app:appendix_reco_APR4} and \ref{app:appendix_reco_SLy} we depict the time-domain reconstructions and their ASD for APR4 and SLy, respectively. The ASD of the signal of the APR4 EOS shows also a noticeable secondary peak at a lower frequency ($\approx 2250$ Hz). This peak, while being present, is not so clearly prominent in the injections and reconstructions of the full merger and post-merger signal (see fourth column in Fig.~\ref{fig:ETTimeWindow}). Both peaks are properly captured by BayesWave up to a distance similar to the one obtained when injecting the full signal. The variability of the highlighted peaks is a sign that different frequencies are present, at different times, on the post merger signal. The number of wavelets used for the reconstructions are displayed in Fig.~\ref{app:appendix_wavelets}, in which the histograms show, as in Fig.~\ref{fig:wavelets}, the number of iterations that use a certain number of wavelets. Since in this case \textsc{BayesWave} only reconstructs the part of the signal corresponding to the inertial-mode emission, the number of wavelets employed for a distance of 10 Mpc is low. 

Fig.~\ref{app:appendix_fpeak} shows the frequency peaks from the ASD of the recovered signals. The larger uncertainty in the case of the APR4 EOS is due to the secondary peak that arises in those injections. The peak from the SLy EOS signal is very well recovered with small uncertainty up to some distance. Even in the case of injecting the part of the signal corresponding to the inertial-mode emission we obtain similar results to the case in which we injected the full post-merger signal. No improvements are obtained and the peak frequency is well recovered up  to a distance of $\approx 12$ Mpc. The overlap between the injected and reconstructed waveforms is depicted in Fig.~\ref{app:appendix_overlap}. As expected, there is a good agreement with the recovery of the frequency peaks. The overlap drops below 0.5 at $d\approx 12$ Mpc, the largest distance at which the peak is recovered with ET.  From these results we conclude that \textsc{BayesWave} yields no difference between reconstructing the full waveform with an early stage in which the signal is much larger or reconstructing only the fraction of the post-merger signal associated with the emission of the inertial modes.

\bibliography{draft}

\end{document}